%% file: main.tex
\RequirePackage{fix-cm}
\documentclass[smallextended]{svjour3}       
\smartqed  
\usepackage{wrapfig}
\usepackage{graphicx}%
\usepackage{multirow}%
\usepackage{array}
\usepackage{fontawesome}
\usepackage{tikz}
\usepackage{tabularx}

\usepackage{appendix}%
\usepackage{xcolor}%
\usepackage{textcomp}%
\usepackage{manyfoot}%
\usepackage{booktabs}%
\usepackage{amsmath}
\usepackage{amssymb}
\usepackage{listings}%
\usepackage{fancyhdr}
\usepackage{fancybox}
\usepackage[utf8]{inputenc}
\usepackage{xcolor}
\usepackage{xspace}
\usepackage{xcolor}
\usepackage{adjustbox}
\usepackage{graphicx}
\usepackage{listing}
\usepackage{cite}
\usepackage{tikz}
\usepackage{bbding}
\usepackage{multirow,tabularx}
\usepackage{enumitem}
\usepackage{caption}
\setlist[itemize]{noitemsep, topsep=0pt, leftmargin=0.1in}
\newenvironment{code}{\captionsetup{type=listing}}{}
\definecolor{dkgreen}{rgb}{0,0.6,0}
\definecolor{gray}{rgb}{0.5,0.5,0.5}
\definecolor{mauve}{rgb}{0.58,0,0.82}
\usepackage{soul}
\usepackage{enumitem}
\usepackage{makecell}

\newcommand{\revision}[1]{{\color{black} #1}}
\newcommand{\revisionM}[1]{{\color{black} #1}}

\newcommand{\apikey}{checked-in secret\xspace}
\newcommand{\apikeys}{checked-in secrets\xspace}
\newcommand{\androidapps}{Android apps\xspace}
\newcommand{\androidapp}{Android app\xspace}
\newcommand{\totalN}{2,142\xspace}

\usepackage{enumerate}
\usepackage{enumitem}
\usepackage{url}
\usepackage{rotating}
\fancypagestyle{plain}{
    \fancyhf{}
    \fancyhead[LE,RO]{\thepage}
}
\begin{document}

\title{Automatically Detecting Checked-In Secrets in Android Apps: How Far Are We?
}
\titlerunning{Automatically Detecting...}
\author{Kevin Li         \and
        Lin Ling     \and
        Jinqiu Yang \and
        Lili Wei
}

\institute{Kevin Li \at
              Department of Electrical and Computer Engineering, McGill University, Canada
              \\
              \email{kevin.li3@mail.mcgill.ca}        
           \and
           Lin Ling \at
              Department of Computer Science and Software Engineering, Concordia University, Canada
              \\
              \email{lin.ling@mail.concordia.ca}
           \and
           Jinqiu Yang \at
              Department of Computer Science and Software Engineering, Concordia University, Canada
              \\
              \email{jinqiu.yang@concordia.ca}
              \and
            Corresponding author - Lili Wei \at
              Department of Electrical and Computer Engineering, McGill University, Canada
              \\
              \email{lili.wei@mcgill.ca}  
}

\date{Author pre-print copy.}

\maketitle
\input{abstract}
\input{introduction}

\input{background}

\input{RQ1}
\input{RQ2}

\input{RQ3}

\input{discussion}

\input{threat}

\input{related}
\input{conclusion}

\section*{Disclosures and declarations}
\textbf{Conflict of Interest.}
The authors declared that they have no conflict of interest.

\noindent \textbf{Data availability.}
Our code and dataset are available on our project website~\cite{data}. For the dataset, we provided sanitized code snippets and data flow of the detected \apikeys. Due to the sensitiveness of the dataset, we did not provide the original \androidapps we used in the experiment.
The \androidapps can be shared to selected researchers and developers upon request.

\noindent \textbf{Funding.}
This work was supported by Fonds de recherche du Québec(Grant No.2024-NOVA-346499)~\cite{nova}, Natural Sciences and Engineering Research Council of Canada Discovery Grant(Grant No. RGCPIN-2022-03744 and Grant No. DGECR-2022-00378) and Fonds de recherche du Québec-secteur Nature et technologies(Grant No.363482~\cite{frqntNewAca}).

\noindent \textbf{Ethical Approval.}
Not Applicable.

\noindent \textbf{Informed Consent.}
Not Applicable.

\noindent \textbf{Clinical Trial Number.}
Not Applicable.

\noindent \textbf{Author Contribution.}
Kevin Li - Conceptualization, Methodology, Data collection, Manual Validation, Formal analysis, Literature Review, Writing.
Lin Ling - Data collection, Manual Validation, Literature Review, Writing - review \& editing. Jinqiu Yang -  Conceptualization, Methodology, Writing review \& editing, Supervision. Lili Wei - Conceptualization, Methodology, Writing review \& editing, Supervision.

\bibliographystyle{IEEEtran}
\newpage
\bibliography{reference}

\end{document}

%% file: abstract.tex
\begin{abstract}

Mobile apps are predominantly integrated with cloud services to benefit from enhanced functionalities.
Adopting authentication using secrets such as API keys is crucial to ensure secure mobile-cloud interactions.
However, developers often overlook the proper storage of such secrets, opting to put them directly into their projects. 
These secrets are checked into the projects and can be easily extracted and exploited by malicious adversaries.
While many researchers investigated the issue of \apikey in open-source projects, there is a notable research gap concerning \apikeys in \androidapps deployed on platforms such as Google Play Store. 
Unlike open-source projects, the lack of direct access to the source code and the presence of obfuscation complicates the \apikey detection for \androidapps. 
This motivates us to conduct an empirical analysis to measure and compare the performance of different \apikey detection tools on \androidapps.

\revision{We conducted a systematic literature review to identify existing \apikey detection tools and then analyzed how these tools could be adapted and applied to the context of \androidapps.} Then, we evaluated three representative tools: Three-Layer Filter, LeakScope, and PassFinder on 5,135 \androidapps, comparing their performance and analyzing their limitations.
Our experiment reveals 2,142 \apikeys affecting 2,115 \androidapps and also discloses that the current \apikey detection techniques suffer from key limitations.
While all the tools achieved high precision, each of them can miss a significant number of \apikeys in \androidapps.

Nevertheless, we observed that the tools are complimentary, suggesting the possibility of developing a more effective \apikey detection tool by combining their insights. Additionally, we outline potential directions for future research, supported by preliminary experimental results.

\end{abstract}

%% file: introduction.tex
\section{Introduction}
Cloud computing is one of the most prevalent technologies in the modern software industry~\cite{cstat}.
Cloud services typically require clients to use secrets (e.g., API keys or API secrets) for authentication and secure access. \revision{In this study, we focus on such cloud API keys that are hardcoded in Android applications as prior work has shown they are the most prevalent type of leaked secret in practice~\cite{twitterregex}.}
Because, in practice, developers often neglect the protection of the secrets and check them into their projects~\cite{basak2023challenges}.
For instance, the usage of hard-coded secrets is one of the 25 most dangerous software weaknesses~\cite{CWE}. Such \apikeys can be exploited once extracted from the released software projects~\cite{798}.
As an example, malicious users can use the extracted secrets to gain access to sensitive information.
In 2017, the API keys of Uber company were mistakenly uploaded to GitHub, resulting in a leakage of information of 50,000 Uber drivers~\cite{uber}.

Most of the existing studies focused on analyzing the \apikey problem in code-sharing platforms such as Github or Gitlab. For example, Meli et al.~\cite{meli} designed a framework based on regular expressions that can identify potential \apikeys within a GitHub repository. Feng et al.~\cite{passfinder} implemented a machine learning tool, PassFinder, to predict whether a string is a \apikey based on its contextual surroundings and intrinsic value.
There also exist a number of community tools that can detect \apikeys in code-sharing platforms~\cite{trufflehog,ggshield,gitleaks,repo-supervisor, whisper, gsh}. Other than developing a \apikey detection tool, some studies~\cite{basak2023challenges,rahman,krause2022committed} conducted empirical analyses to investigate the \apikey issue on the code-sharing platform.
 
Meanwhile, there is a noticeable research gap of the \apikey problem on \androidapps deployed on the Google Play Store.
\androidapps are also known to be susceptible to the problem of \apikeys~\cite{basak2023challenges}.
Checked-in secrets can be extracted from the deployed \androidapps through unpacking and reverse engineering~\cite{extractTutorial}.
Nevertheless, few studies have investigated this problem.
Zuo et al.~\cite{LeakScope} proposed the only tool, LeakScope, for \apikey detection in \androidapps. 
Their study revealed that over 15,000 \androidapps have the \apikey problem. However, their study is limited to three cloud providers, Google, Amazon, and Microsoft and requires substantial time and computational resources to expand its detection capabilities and analyze Android apps. \revision{While this research aims to broaden the scope of \apikey detection, it also highlights a critical challenge: many Android applications available on app stores are only distributed as compiled APKs and are frequently obfuscated~\cite{obfuscationprac}. This combination restricts access to high-level code constructs and undermines techniques that depend on contextual or semantic information, such as machine-learning-based approaches. Additionally, since APKs are also the only artifacts accessible to potential attackers, evaluating secret detection in this format realistically simulates how secrets may be retrieved from closed-source apps. Furthermore, because the Google Play Store is the official marketplace and central distribution channel in the Android ecosystem, studying apps in this setting is both practically significant and security-relevant. These characteristics make Android apps particularly difficult to analyze and highlight why dedicated studies of secret detection in this setting are necessary.}

In this paper, we aim to fill the existing research gap by evaluating the performance of state-of-the-art \apikey detection tools, \revision{each representing a distinct detection approach}, when applied to Android apps. It also provides insights for the future development of Android \apikey detectors.
We evaluate the performance of Android-specific \apikey detector, LeakScope, as well as adapting source-code-level \apikey detectors designed for code-sharing platforms. 
The source-code-level \apikey detectors can potentially be adapted to \androidapps since the source code of the apps can be recovered through decompiling or reverse engineering.
However, it is still challenging to directly leverage the source-code level \apikey detectors in \androidapps due to the wide adoption of obfuscation.
Obfuscation obscures the lexical information of the code that is often needed by the source-code-level \apikey detectors. We also want to study the impact of obfuscation on these source-code-level checked-in secrets detectors.

In summary, we want to answer the following research questions in this work:

\begin{itemize}
    \item \textbf{RQ1: (Survey of existing tools)} \textit{What are the existing \apikey detection approaches? Which tools can be adapted to \androidapps?}

    \item \textbf{RQ2: (Evaluation and comparison)} \textit{How does each secret detection tool perform when adapted to \androidapps, and how do these tools compare against one another?}

    \item \textbf{RQ3: (Implications)} \textit{What are the limitations of the evaluated tools? How can we improve checked-in secret detection for Android apps?}

\end{itemize}

To achieve the proposed research objectives, we conducted a systematic literature review to identify the available academic and community \apikey detection tools. we discovered seven academic \apikey detection tools and six community tools. we further categorized the identified tools into three categories based on their methodology: Intrinsic value analysis, static analysis and machine learning-based analysis. we selected the framework proposed by Meli et al. (intrinsic value) ~\cite{meli}, LeakScope (static analysis) ~\cite{LeakScope}, and PassFinder (machine learning) ~\cite{passfinder} as the representative tool for each category based on their performance, availability, and adaptability.

We then made necessary modifications and extensions for each tool to be adapted to \androidapps and launched them on 5,135 \androidapps deployed on the Google Play Store. Overall, \totalN valid \apikeys were detected in 2,115 \androidapps. We then compared the performance of each tool based on the number of detected \apikeys and produced false positives/negatives. Our findings indicate that, although all the selected tools limited the presence of false positives, they still have significant limitations when applied to \androidapps. 

We observed that all the evaluated tools have a significant number of false negatives.
The detection capability of both Meli et al.'s work and LeakScope is largely constrained by pre-defined inputs to specify the detectable types of \apikeys (i.e., regular expressions for Meli et al.'s work and cloud API signatures for LeakScope).
In contrast, PassFinder's detection capability is not limited by pre-defined inputs, but the presence of obfuscation in \androidapps greatly reduces its performance.

Despite the limitations of each tool, we observe that these approaches can complement each other. For instance, the data flow analysis of \apikeys performed by LeakScope could assist Meli et al.'s work in extracting \apikeys that would otherwise go undetected. The methodology utilized by Meli et al. can enhance \apikey prediction accuracy as well.
Additionally, analyzing string groups within methods containing \apikeys could be a more effective indicator of the presence of \apikeys, as obfuscation does not alter these strings.

To summarize, we make the following major contributions in this paper:
\begin{itemize}
\item To the best of our knowledge, we are the first to conduct an empirical study of the \apikey issues in \androidapps deployed on the Google Play Store.
\item We adapted existing \apikey detection tools that are originally designed for open-source projects to \androidapps deployed on the Google Play Store.
\item We evaluated and compared the performance of the existing \apikey detection tools. We analyzed the limitations of existing \apikey detection tools and proposed future research directions for developing more effective \apikey detection tools tailored for \androidapps.
\item \revision{ We provided a benchmark dataset consisting of the sanitized code snippets of the validated \apikeys in \androidapps   and the complete dataset can be access upon request.}
\end{itemize}

%% file: background.tex
\section{Background}

\subsection{Checked-in Secrets}

Cloud service providers deploy various methods for authenticating processes to minimize security risks~\cite{lu2011accessing}. Using the secrets to authenticate cloud service is the most common approach.
When incorporating cloud services into an application, cloud service providers require developers to generate secrets. 

These secrets allow the cloud service providers to identify which project is calling the cloud API, verify whether it has access to certain resources, and manage billing functions.

Leaking the sensitive cloud service secrets to the public can result in severe security consequences for both cloud service providers and applications.
Access to the \apikeys allows anyone to utilize the allocated resources and functionalities.
It can lead to (1) data leakage if the \apikey is used to authenticate access to application data, (2) data corruption if the \apikey allows for data manipulation, (3) monetary losses if it creates costs for invoking the concerned APIs, and (4) denial of service (DoS) if there is a limit for the API invocation.

\begin{code}

\begin{lstlisting}[language=Java,
    numbers=left,
    numbersep=5pt,
    frame=single,
    tabsize=2,
    caption={Code snippet with one \apikey. It is extracted from an \androidapp in our experiments.},
    label={lst:APKSecretExample}
]
task.addOnSuccessListener(listener)
context = r0.getApplicationContext()>()
r7.ggh(context, "AIza------")
r0.setupLifecycleListener()
\end{lstlisting}
    \end{code}

As an example, Listing \ref{lst:APKSecretExample} is a code snippet extracted from one \androidapp in our experiment dataset. The string at line 3 that starts with \textquotedblleft Alza\textquotedblright is a Google API key, which could be used to grant access to Google Map services. This secret can be used to perform a DoS attack since the API has a limit of requests per minute or cause financial losses since it costs \$0.005 per request.

\subsection{APK Files}
Unlike the open-source \androidapp projects on GitHub, the \androidapps deployed on the Google Play Store are distributed as APK (Android Package Kit) files. An APK file is essentially an archive file that contains all the information needed for an \androidapp to be successfully executed on a device. When a user wants to download an \androidapp from platforms such as the Google Play Store, the platforms will automatically download the APK files for the users. Several automatic tools exist, such as ApkTool~\cite{apktool}, dex2jar~\cite{dex2jar}, Soot~\cite{soot}, and Mobsf~\cite{mobsf}, that can perform reverse engineering on an APK file. However, these reverse engineering tools all have limitations. They may not accurately recover the source code from APK files. Sometimes, they might even miss some classes while analyzing the APK files. Furthermore, APK files commonly adopt obfuscation techniques to enhance the application's security~\cite{obfuscation}.
Obfuscation obscures the code of the \androidapps by altering the names and hierarchies of code artifacts.
For example, in line 3 of Listing \ref{lst:APKSecretExample}, the method \texttt{ggh} was originally a method called \textquotedblleft initialize()" provided by Google to send an API request. Those factors are the main challenges that need to be addressed for the tools focused on \androidapps. The effectiveness of detection tools relying on code context is significantly reduced by obfuscation, which makes variables and class names meaningless.

%% file: RQ1.tex
\section{RQ1: Survey of Existing Tools}
The checked-in secrets issue is a long-studied subject in the software development field. There exist many tools that aim to detect \apikey. In RQ1, we aim to find these available academic and community \apikey detection tools and analyze whether they can be adopted on \androidapps through a detailed literature review process.
\subsection{RQ1-Methodology: Literature Review}
We conducted a literature review to gather a comprehensive list of \apikey detection tools/methodologies. The search query is composed of two parts, as shown in Table~\ref{tab:keywords}:  Part 1 includes terms related to secret types (e.g., API key, secret, password), while Part 2 focuses on detection methodologies.
\begin{table}[t]
\caption{Search String Group}
\centering
\begin{tabular}{c|c}
\toprule

\textbf{Part 1} & \textbf{Part 2} \\ \midrule
(API) key, secret, credential, password & detection, leakage, exposure \\ \bottomrule

\end{tabular}

\label{tab:keywords}
\end{table}

Each search query combined Part 1 and Part 2, yielding 12 queries (4 × 3), such as \textquotedblleft API key detection." We conducted the initial search in Google Scholar, IEEE Xplore, and the ACM Digital Library in January 2024.

Each query returns a large number of search results, including irrelevant ones. For example, some queries, such as API key detection, returned over 10,000 results on the ACM digital library.
\revision{
Manually analyzing and selecting relevant papers from such a large pool is impractical. Therefore, for each search platform, we chose to sort the results by relevance, as this is typically the default and recommended ordering used by digital libraries to display the most pertinent publications first. This approach increases the likelihood of capturing high-quality and topically appropriate papers early in the result set. And if a query returned more than 500 results, only the top 500 most relevant results were analyzed.} In total, 5,500 papers were selected for the initial filtering process.

During the initial filtering process, \revision{one of the authors manually reviewed the titles of all 5,500 papers. Any paper with a title not related to key terms such as API, security, secret detection, or privacy leakage was excluded. This manual screening step resulted in the selection of 78 papers for further analysis.}
 Subsequently, we conducted a more thorough review of these 78 papers to identify those most relevant to our subject. In this stage, two authors read the abstracts (and, if necessary, the full content) of each paper to determine its relevance. If there was any disagreement, a discussion was held to reach a consensus. \revision{For example, the paper \textit{\textquotedblleft Your Code Secret Belongs to Me: Neural Code Completion Tools Can Memorize Hard-Coded Credentials"}~\cite{secrectllm} was initially included after the title-based filtering, but later excluded during this stage as it focuses on credential memorization in neural code completion tools rather than on detection techniques.}
 In total, 19 papers were selected after this step.

We also conducted a snowballing process, a strategy that is commonly adopted in other literature review research~\cite{Liu_2022}. This ensures we do not miss any important papers on the \apikey. For each paper that we initially selected, we reviewed references from the selected papers as well as papers that cited them. This process continued until no new relevant papers were identified.
\begin{figure}[t]

\centering
\includegraphics[width=\linewidth]{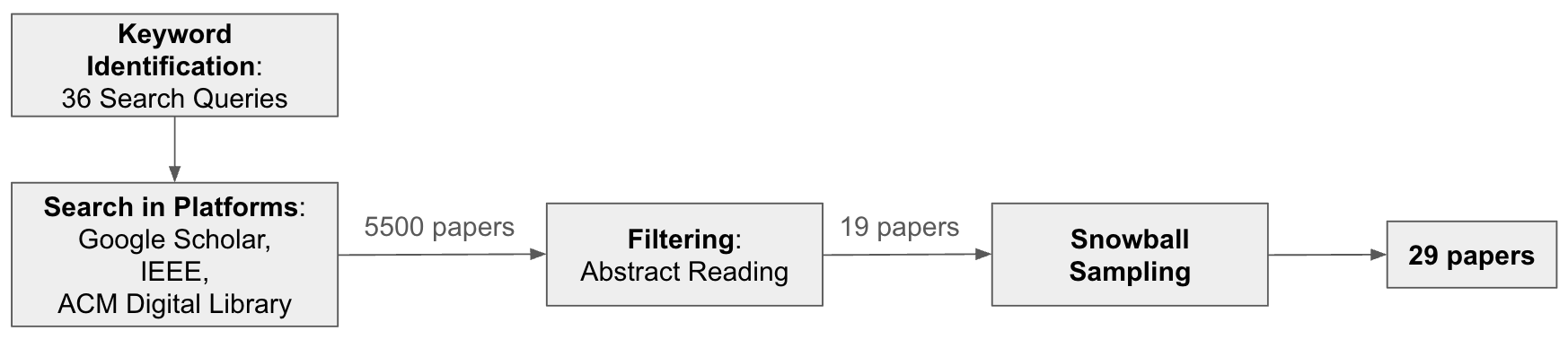}

\caption{Literature Review Process.}
\label{process_l}
\end{figure}
Eventually, we found 29 academic papers that are related to \apikey detection. The literature review process is illustrated in Figure \ref{process_l}.

For community tools, we used the six community tools selected by S. K. Basak et al. ~\cite{emstudy} as they conducted a thorough selection process to perform an analysis of \apikey detection tools on the open-source repositories.

\subsection{RQ1-Result}
Among the 29 papers, seven of them developed a \apikey detection tool or methodology of our interest, while others are either empirical studies on \apikey issue and management or only specific to platforms such as DockerHub. Overall, a total of 13 \apikeys detection tools, when combined with community-developed tools, were collected. 

We categorized the \apikey detection tools into three groups: \textbf{Intrinsic Value Analysis}, \textbf{Static Analysis}, and \textbf{Machine-Learning-Based Analysis} according to their detection approach. Table \ref{tab:tool_summary} presents the list of description of all the identified \apikey detection tools and their corresponding categorization. Overall, there are seven Intrinsic value analysis tools, one static analysis tool, and five Machine Learning based analysis tools.

\begin{table}[t]
\centering
\caption{Categorization, Availability, and Description of Tools}
\label{tab:tool_summary}
\begin{adjustbox}{max width=\textwidth}
\begin{tabular}{c|c|c|c|p{6.5cm}}
\toprule
\textbf{Category} & \textbf{Tool} & \textbf{Selected} & \textbf{Availability} & \textbf{Description} \\
\midrule
\multirow{7}{*}{Intrinsic Value Analysis} 
& Meli et al. [7] & \Checkmark & \Checkmark & A Three-Layer Filter combining an entropy-based filter, a word (dictionary) filter, and a pattern filter \\
& Gitleak [11] & \XSolidBrush & \Checkmark & Detects hard-coded secrets in GitHub through regex-based pattern matching \\
& Repo Supervisor [12] & \XSolidBrush & \Checkmark & Detects hard-coded secrets in supported file types via regex-based pattern matching and entropy calculation \\
& Trufflehog [9] & \XSolidBrush & \Checkmark & Detects hard-coded secrets in GitHub via regex-based pattern matching and entropy calculation \\
& Whispers [13] & \XSolidBrush & \Checkmark & Parses structured files (e.g., JSON, YAML, XML) to extract key–value pairs, then applies regex-based rules \\
& GGshield [10] & \XSolidBrush & \Checkmark & Detects hard-coded secrets in GitHub through regex-based pattern matching and entropy calculation \\
& GitHub Secret Scanner [14] & \XSolidBrush & \Checkmark & Detects hard-coded secrets in GitHub through regex-based pattern matching and entropy calculation \\
\midrule
Static Analysis 
& LeakScope [18] & \Checkmark & \Checkmark & A static analysis tool for Android apps that performs inter-procedural backward data-flow analysis from predefined cloud API calls to extract secrets \\
\midrule
\multirow{5}{*}{Machine Learning Analysis} 
& PassFinder [8] & \Checkmark & \Checkmark & Identifies secrets in code by analyzing intrinsic string characteristics and surrounding code context \\
& Secrethunter [29] & \XSolidBrush & \Checkmark & Uses reinforcement learning to filter unlikely files on GitHub, then applies regex, entropy checks, and dictionary-based filtering \\
& Seagull [30] & \XSolidBrush & \Checkmark & Detects secrets by analyzing code snippets and their data flow \\
& Saha et al. [31] & \XSolidBrush & \XSolidBrush & Uses regex-based candidate extraction followed by analyzing string features, entropy, and file metadata \\
& Lounici et al. [32] & \XSolidBrush & \XSolidBrush & Combines regex-based secret scanning with two ML classifiers—one for file path context and another for code snippet content \\
\bottomrule
\end{tabular}
\end{adjustbox}
\end{table}

\textbf{\textit{Intrinsic value analysis}}.
This type of \apikey detection tools focuses on the intrinsic value of the strings under analysis.
The analysis is generally done using two processes: regular expression matching and entropy calculation.
For regular expression matching, in the cases of certain cloud providers, the generated \apikeys often adhere to the same format. For instance, Google’s checked-in secrets always take the form of “ALza******”. Utilizing regular expression matching enables the extraction of such checked-in secrets from the code. 
For entropy calculation, as \apikeys are all randomly generated strings, the randomness of the characters in a string will be a good indicator of \apikey. These techniques leverage Shannon entropy to give a good estimation of randomness. A randomly generated string will generally have higher entropy than a normal string.

\textbf{\textit{Static analysis.}}
To the best of our knowledge, LeakScope\cite{LeakScope} is the only one tool that extracts \apikeys through static analysis . By providing a location where a \apikey is used, static analysis can calculate the value of the \apikey through data and control flow analysis.

\textbf{\textit{Machine Learning based analysis}}.
This type of \apikey detection tool takes into consideration the fact that the contextual surroundings of \apikeys follow some specific patterns. Factors such as file type, class name, variable name, and code snippet could be used as features to train a machine learning model to predict whether a string is a \apikey.

\subsection{Tool Selection for Further Evaluation}
The \apikeys detection tools within the same category often adopt similar techniques. For example, \apikey detection tools based on intrinsic value analysis typically use regular expression matching, entropy calculation, or a combination of both. \revision{Since our goal is to analyze the detection approaches rather than individual tools, we chose one representative tool from each category for further evaluation in RQ2 and RQ3. Tool selection was guided not only by effectiveness and adaptability, but also by how well each tool represents its respective approach. We made sure that all the selected tools cover key technologies for each category.}

For intrinsic value analysis, we selected the Three-Layer Filter technique proposed by Meli et al.~\cite{meli} \revision{because it comprehensively incorporates all major techniques commonly used in this category. Specifically, (1) intrinsic value analysis typically relies on manually crafted regular expressions, and the Three-Layer Filter provides a well-defined and precise set of such patterns; and (2) in addition to regex-based matching, it also applies entropy filtering, word filtering, and pattern filtering, thereby covering the full range of strategies adopted by other tools in this class.} Furthermore, (3) it has been shown to be one of the most effective tools in this category~\cite{emstudy,meli}.

For the representative Static Analysis tool, we chose LeakScope because it is the only tool designed for Android APK files and is available on GitHub~\cite{LeakScopeGithub}.

For Machine-Learning-Based tools, we selected PassFinder for further evaluation.
PassFinder, SecretHunter, and Seagull are the only publicly available tools. \revision{Among them, PassFinder aligns best with the scope of our study. SecretHunter is specifically designed to address the challenge of missing files in GitHub repositories, which often results from API or bandwidth limitations, making it tightly coupled with the GitHub platform and less applicable to our Android-based context. Seagull, on the other hand, relies heavily on CodeQL~\cite{codeql} for code representation, which is incompatible with our analysis setup as it is designed only for open-source projects on Github. Additionally, both PassFinder and Seagull incorporate two key types of features in their detection process: intrinsic string properties and contextual surroundings. Since these two categories represent the most widely adopted feature sets in machine learning-based secret detection, PassFinder’s design makes it a representative choice for our evaluation. Considering practical factors such as tool availability, compatibility, and feasibility of integration into our experimental setup, we selected PassFinder as the representative machine-learning tool.}

By selecting these representative tools, we ensured that each category of checked-in secret detection methodology was evaluated in the context of \androidapps. Below, we describe the three representative \apikey detection tools in detail.

\noindent\textbf{Three-Layer Filter.} Meli et al.~\cite{meli} is the first to conduct a large-scale analysis on the \apikey problem of the open-source projects on GitHub, covering 13\% of all the public repositories. Their methodology involved initially extracting strings from the repositories through regular expression matching. Subsequently, these candidate strings go through a Three-Layer Filter by evaluating the candidate string's entropy value, pattern, and word.
 First, the authors designed 15 regular expressions for 12 different cloud providers and applied the 15 regular expressions to extract candidate strings from the source code.
Then, the extracted candidate strings will be filtered by three layers. We refer to this technique as Three-Layer Filter for the rest of this paper.
Specifically, the three filers are:

\textit{- Entropy Filter.} The entropy filter first calculates the Shannon Entropy~\cite{shannon} for each string, and within the same regular expression group, a string will fail if its Shannon entropy is three standard deviations away from the mean.

\textit{- Word filter.} Meli et al. also observed that if a string contains an English word, it is nearly impossible for the string to be a \apikey. Therefore, a string will fail if it contains words from a pre-defined set of English words. 

\textit{- Pattern filter.} 
Meli et al. also observed that there are some patterns that are unlikely to appear in \apikeys. Therefore, they proposed that a string should not be considered as \apikey if it contains patterns such as repetitive, ascending, and descending characters.

\noindent\textbf{LeakScope~\cite{LeakScope}} is the first automated tool that aims to detect the data leakage problem caused by \apikey leakage and misconfiguration of secret permission in \androidapps. Their tool is developed based on the key finding that if a project wants to use the services provided by cloud providers, the project would likely call the cloud APIs provided by the cloud providers (i.e., the APIs defined in the cloud SDKs). These cloud APIs generally take one or several \apikeys as part of their parameters. Take the \texttt{CloudStorageAccount.parse(String)} API of Microsoft Azure as an example; this cloud API takes a \apikey as a parameter and returns the account that is connected to this \apikey ~\cite{cloudaccount}; LeakScope collected 32 cloud APIs that take \apikey as parameter from three cloud providers: Microsoft Azure, Amazon AWS, and Google Firebase. To extract the value of the \apikey, LeakScope performs an inter-procedural backward analysis starting from the cloud APIs; along with the backward analysis, LeakScope keeps track of the data flow associated with the parameter of this cloud API. Through this data flow tracking, LeakScope proceeds to compute the value of the parameter. Moreover, LeakScope employs a signature generation mechanism for cloud APIs based on their package structure and content, enabling identification of their usage within the project. This signature empowers LeakScope to operate effectively, even in scenarios where APK files are obfuscated.

\noindent\textbf{PassFinder~\cite{passfinder}}
is a \apikey detection tool utilizing the deep neural network. It conducts an analysis of strings by considering both the intrinsic value of the string itself and its contextual surroundings, particularly encompassing twelve lines of context. PassFinder trains separate models for the intrinsic value and context of the string. The prediction is given by combining the results of these two models.
PassFinder is designed to analyze passwords and API credentials in GitHub repositories, as it is trained with code snippets collected from GitHub repositories. It is also capable of handling different programming languages. 

\revision{Both the Three-Layer Filter and PassFinder were originally designed to operate at the source-code level, where raw code and file structures are directly accessible. In contrast, LeakScope relies on Soot to analyze Java programs by converting them into an intermediate representation known as Jimple. Since our study is based on Android APK files, where the original source code is typically unavailable, we adapted our approach accordingly. Specifically, to enable a fair comparison and apply LeakScope in our setting, we reimplemented the Three-Layer Filter and PassFinder using Soot to process APK files and extract the necessary Jimple representation for analysis.}

%% file: RQ2.tex
\section{RQ2: Comparison of Existing Tools}
In RQ2, we aim to evaluate and compare the performance of each category's selected representative \apikey detection tools on \androidapps.

\subsection{RQ2-Methodology}
We aim to evaluate and compare the performance of the selected tools based on the following criteria:
\revision{
\begin{itemize}[leftmargin=*]
    \item \textbf{RQ 2.1: Number of detected \apikeys:} How many \apikeys can be detected by each of the selected tools?

    \item \textbf{RQ 2.2: False positives:} How many \apikeys reported by each tool are false positives? 
    \item \textbf{RQ 2.3: False negatives:} 
    How many false negatives are produced by each tool?
\end{itemize}}

\subsubsection{Dataset}
We collected a dataset of \androidapps for evaluation, using Raccoon APK Downloader~\cite{raccon} to retrieve apps from the Google Play Store. Based on AppBrain~\cite{appBrain} statistics, we selected the top 200 apps in the top 30 categories like Business, Movies, and Music. After filtering out apps that could not be processed by Soot, we obtained a total of 5,135 APK files.

The experiment was conducted on a High-Performance Computing cluster that consisted of 24 compute nodes with 32 cores and 512 GB of memory and approximately 10 TB of volatile-scratch disk space. We utilized a total of 32 CPUs for 16 jobs, with each job using 2 CPUs running concurrently.

\subsubsection{Tool Adaptation}
None of the selected tools is ready to use for the experiments. The Three-Layer Filter did not provide the necessary code, LeakScope did not provide the cloud API signatures, and PassFinder did not provide the pre-trained models. Thus, some implementation efforts were necessary for the experiments. 
Additionally, unlike PassFinder, whose detection capability is not strictly constrained, LeakScope and the Three-Layer Filter can only detect the \apikeys that are used as inputs in the pre-defined cloud APIs or adhere to the pre-defined regular expressions. 
Therefore, expanding the list of cloud APIs and regular expressions was necessary to equalize the detection capabilities of LeakScope and the Three-Layer Filter and conduct a fair comparison. 

In this section, we detail the adaptations we made for all the tools.

\textbf{Adaptation for Three-Layer Filter.}
\revision{Although the source code for the filtering process of the Three-Layer Filter is not provided by the authors, we implemented the necessary components based on the explanations in the paper~\cite{meli}, which were clear and sufficient to guide our reimplementation.}
For the string extraction, since the source code is not directly accessible in APK files, we used Soot~\cite{soot} \revision{to identifying all the string constants within the APK files, aligning with the original focus of the technique. }

The authors provide a set of regular expressions to identify potential API keys. \revision{In our reimplementation, we used the same set of regular expressions.} However, we observed that in their experiment, the filtering process excluded only 2\% of the strings, indicating that their regular expressions were highly precise and rarely mismatched actual API keys. As a result, the filtering process had minimal impact on reducing false positives. 

However, it is challenging to obtain precise regular expressions for all the categories of \apikeys in practice. It is more common for cloud providers to adopt loose regular expressions. For example, \apikeys of Twitter and Facebook do not exhibit a strong pattern. Any combination of letters and numbers that fall within a specific range of lengths can match the regular expressions. \revision{To fully evaluate the effectiveness of the filtering mechanism, we introduced two additional, more loose regular expressions from a previous study~\cite{twitterregex} to simulate a scenario where the initial candidate set may contain a higher number of false positives. This allowed us to better assess the contribution of the subsequent filtering process.}
 
Therefore, we used two more regular expressions derived by a previous study~\cite{twitterregex}, as shown in Table \ref{tab_reg_2}. We chose Twitter client ID and secret because the corresponding \apikeys do not exhibit a strong pattern; Twitter APIs are commonly used in \androidapps.

\begin{table}[h]
\centering
\caption{Newly proposed regular expressions}
\begin{tabular}{c|c}
\toprule
\textbf{Cloud providers} & \textbf{Regular expression}\\
\midrule
Twitter Client ID   & [0-9a-zA-Z]\{18,25\}\\ \hline
Twitter Client secret  & [0-9a-zA-Z]\{40,50\} \\ \bottomrule

\end{tabular}

\label{tab_reg_2}

\end{table}

\textbf{Adaptation for LeakScope.}
In order to use LeakScope to extract the \apikeys from an APK file, we have to provide the complete signature of the cloud API and specify which API parameter holds \apikeys to LeakScope. The authors of LeakScope provided the names of 32 cloud APIs from Amazon AWS, Microsoft Azure, and Google Firebase, but they did not provide the complete signatures of the cloud APIs. \revision{To address this, we consulted the official documentation of each corresponding cloud API to collect the full and accurate signatures of these APIs.} The full list of signatures can be found on our website~\cite{data}. 

However, the provided cloud APIs are limited to three cloud providers. The number of supported cloud providers is very limited compared to the Three-Layer Filter.
To perform a more in-depth and fair analysis, we aim to collect more cloud APIs that can be analysis starting points for LeakScope, i.e., the cloud APIs that take \apikey as a parameter.
Given the abundance of cloud providers in the contemporary market, attempting to collect all the cloud APIs that take \apikey as a parameter from every single provider would be infeasible, especially since some cloud providers may not offer comprehensive API documentation. 
Therefore, instead of solely relying on the API documentation provided by every cloud provider, our cloud API collection strategy also takes into account the potential cloud API extracted from our dataset.

We first collected 15 cloud providers, including those covered by the Three-Layer Filter but not by LeakScope, as well as the top 13 cloud service providers in 2024~\cite{topCP}.
Then, we read the available API documentation of the collected cloud providers to identify the cloud APIs that use secrets as part of their parameters.
To mitigate the risk of missing some relevant API documentation, we also adopted an additional step.
We collected information of all the methods that are called at least once in the \androidapps of the dataset.
We then selected all the methods that contain the names of the collected cloud providers.
By analyzing the context and implementation (API documentation if available) of these methods, we determined whether they are our targeted cloud APIs. For example, a constructor, 
\texttt{OSSStsTokenCredentialProvider(String, String, String)} was not originally found by reading the API documentation of Alibaba, but as its package name contains \textit{Alibaba}, we were able to identify this cloud API.

Overall, we collected 13 new cloud APIs from five additional providers. For the remaining 10 cloud providers, we failed to recover any cloud APIs that take \apikeys as input.
This results in a total of 48 cloud APIs from 10 cloud providers. The authors of LeakScope provided 32 cloud APIs from Amazon, Google, and Microsoft. By analyzing API documentation, we identified 10 additional cloud APIs from Alibaba, Facebook, Redhat, Twitter, and Stripe. Furthermore, we collected 6 new cloud APIs from Alibaba, Facebook, Paypal, and IBM Cloud.

We acknowledge that our collection process cannot capture every cloud API using \apikeys as parameters, but our selection is representative of the typical APIs used in Android apps. Overall, they greatly enhanced the detection capability of LeakScope.

\textbf{Adaptation for PassFinder.}
We contacted the author to request the pre-trained models, but they indicated that the project is not currently being maintained, but the necessary dataset and instructions are provided~\cite{passFdATA}.
As a result, we followed the exact instructions provided by the author to train the two models provided by PassFinder. 

For the password model, we leveraged the original training set provided by the authors~\cite{passFdATA}.
For the context model, the original training set contains 12,000 code snippets for the ten most popular programming languages on GitHub.
For each language, there are 600 snippets with \apikeys and 600 snippets without \apikeys.

Once the password and context models were trained, we utilized the password model to identify potential \apikeys within our dataset. Subsequently, we employed Soot~\cite{soot} to extract the contextual surroundings of the identified \apikeys, \revision{specifically capturing six instructions before and six instructions after each occurrence within the same method body, which is the same as the definition of contextual surroundings in the paper of PassFinder~\cite{passfinder}. These surrounding instructions were then used as input to the context model for detection.} However, instead of using source code, Soot will output the contents of APK files in Jimple format, an intermediate representation of the Java language. Jimple representations are very similar to Java syntax, so we decided to remain with the Jimple format for the extracted contextual surroundings.

\begin{table*}[h]
\centering
\caption{The type of \apikeys that each tool can detect}

\begin{tabular}{cc}
\toprule
\textbf{Tools}& \textbf{Detection capability (Cloud providers)}  \\\midrule
The Three-Layer Filter & \makecell{Amazon, Microsoft, Facebook, Twitter client ID, \\Twitter client secret, Twitter access token, \\Google API key, Stripe, Square, Picatic,\\  PayPal Braintree, Twilio, MailGun, MailChimp}  \\ \hline
LeakScope & \makecell{Amazon, Microsoft, Facebook, Twitter client ID,\\ Twitter client secret, Google API key, Stripe,\\ Square Redhat, IBM cloud}  \\ \hline
PassFinder & Not strictly constrained  \\

 \bottomrule

\end{tabular}
\label{tab:d}
\end{table*}

\subsubsection{Comparison of the Detection Capability of Each Tool} \label{sssec:dc}
After making the necessary adaptation for each tool, the detection capability of each tool is listed in Table \ref{tab:d}. Overall, we extended LeakScope's detection capability to align with the Three-Layer Filter. However, it is infeasible to identify a comprehensive set of cloud API signatures for LeakScope that match all the regular expressions used by the Three-Layer Filter and vice versa. This difficulty arises because some cloud providers do not issue secrets that follow specific patterns that can be precisely captured by regular expressions (e.g., Alibaba, Redhat, and IBM). Furthermore, we failed to find the cloud APIs of the cloud providers that the Three-Layer Filter covered  (e.g., MailGun, MailChimp, and Picatic). Although the detection capability of LeakScope and the Three-Layer Filter is not entirely equalized, our extension ensured a more meaningful comparison among the tools.

To enable a fair comparison, we present detailed detection results for each cloud provider in the following sections. \revision{For the Three-Layer Filter, each detected secret is mapped to its corresponding cloud provider based on the regular expression used for detection. In the case of LeakScope, the mapping is derived from the cloud provider associated with the matched API in our collected cloud API list. For PassFinder, the evaluation dataset includes some secrets that were also detected by the other two tools. In such cases, we adopted the labels generated by those tools as the ground truth. For the remaining cases, we manually mapped the secrets to cloud providers by analyzing the surrounding context. While the detection capabilities of the tools are not perfectly aligned, these mapping strategies allow for a more meaningful and fair comparison. In the following sections, we present the detailed detection results per cloud provider.}

\subsubsection{Checked-in Secret Validation and Ethical Considerations.}\label{sssec:ethics}
A key challenge in evaluating \apikey detection tools is the lack of publicly available ground truth datasets for \apikeys in \androidapps. To the best of our knowledge, no benchmark dataset currently exists. While full validation of an \apikey would ideally involve testing it against the corresponding cloud service, doing so may violate terms of service or raise ethical and legal concerns.
\revision{
To address this, we adopted a manual analysis procedure commonly used in prior \apikey research~\cite{emstudy,secretbench}. Specifically, a secret was considered valid if it met the following criteria: (1) it matched the expected format or structure defined by known cloud API key patterns (e.g., prefixes such as AIza for Google), and (2) it appeared in a relevant context, such as within a function call to a known cloud API or near related SDK usage. For instance, in Listing~\ref{lst:APKSecretExample1}, the presence of functions from the Google Cloud Library helped confirm the validity of the suspected \apikey.}

The manual analysis was conducted independently by two authors. Each author reviewed the format and context of the detected string. In cases of disagreement, both reviewers discussed their reasoning, referenced available documentation when needed, and reached consensus through mutual deliberation. This structured review process allowed us to evaluate suspected \apikeys consistently while avoiding direct interaction with external services.

\begin{code}
\begin{lstlisting}[language=Java,
    numbers=left,
    numbersep=5pt,
    frame=single,
    tabsize=2,
    caption={Code snippet with one \apikey\ and its context. It is extracted from an \androidapp\ in our experiments.},
    label={lst:APKSecretExample1}
]
GoogleApiClient GoogleClient = x.GoogleClient;
d.b0.c.l.c(GoogleClient);
mGoogleApiClient.disconnect();
GoogleSignInOptions.Builder builder = new Builder;
builder = builder.requestIdToken("");
builder = builder.requestEmail();
\end{lstlisting}
\end{code}

To complement our manual analysis, we validated the functionality of \apikeys identified as Google API keys by sending test requests to the Google Maps service, in order to verify whether the detected secrets were truly usable. We selected this service because the requests do not expose any sensitive user information and allow safe testing without accessing private data. \revision{This validation step provides a practical snapshot of the potential real-world impact of leaked secrets by demonstrating that many of them are still active and exploitable.} To avoid any significant financial impact, each \apikey was used only once. Additionally, for every valid \apikey we detected, we notified the developers of the corresponding \androidapps.

\subsection{RQ2.1 Number of Detected Checked-in Secrets}
\subsubsection{RQ2.1-Methodology} \label{sssec: 2_1_M}
In this section, we aim to evaluate the effectiveness of the tools by evaluating the number of \apikeys that can be detected by each tool. As the detection capability of each tool is not identical, we evaluated the number of detected \apikeys identified by each tool across various cloud providers to provide a fairer comparison among the tools.
The detailed result of the number of detected \apikeys is presented in Table \ref{tab:filter}. Note that the detected \apikeys in this table are not necessarily true positives. In this research question, we demonstrate the overall results reported by each tool. We conduct an analysis on the false positives of the tools in Section~\ref{ssec:fp}.

\begin{table*}
\centering
\caption{Number of detected \apikeys corresponding to the cloud providers of the three selected tools, the detected \apikeys are not verified. NA signifies that the \apikeys of the corresponding cloud providers are not in the detection capability of the tool.}
\label{tab:filter}
\begin{tabular}{lrrr|r}
\toprule
\tiny \textbf{Cloud Providers} & 
 \tiny \makecell{\textbf{Three-Layer}\\ \textbf{Filter}} &  \tiny \textbf{LeakScope} & \tiny \makecell{\textbf{PassFinder}}& \tiny \makecell{\textbf{Total \# of Unique } \\ \textbf{Checked-in Secrets} \\ \textbf{per Cloud Provider}}\\
\midrule
Amazon&0&5&0&5\\
Microsoft&0&0&0&0\\
Facebook&0&10&0&10\\
Twitter Access Token&0&NA&0&0\\
Twitter Client ID/Secret&88,172&10&0&88,182\\
Google&664&1,816&0&2,117\\
Stripe&0&0&0&0\\
Square&0&0&0&0\\
Picatic&0&NA&0&0\\
PayPal Braintree&0&0&0&0\\
Twilio&2&NA&0&2\\
MailGun&0&NA&0&0\\
MailChimp&0&NA&0&0\\
Redhat&NA&0&0&0\\
IBM Cloud&NA&0&0&0\\
Alibaba&NA&0&0&0\\ \hline

\tiny \noindent \makecell{\textbf{Total \# of Checked-in} \\ \textbf{Secrets of Each Tool}} &88,838&1841&0&90,316 \\
\bottomrule

\end{tabular}
\end{table*} 
\subsubsection{RQ2.1-Results}
Table \ref{tab:filter} lists the number of detected \apikeys of each tool. In total, 90,316 unique \apikeys were identified by the three tools. The Three-Layer Filter successfully identified 88,838 \apikeys. LeakScope successfully identified 1,841 \apikeys. For PassFinder, the password model successfully identified 11,343 \apikeys, but the context model failed to identify any \apikeys. 

The Three-Layer Filter detected the highest number of unique \apikeys, \revision{while PassFinder failed to detect any. PassFinder's limited effectiveness can be partially attributed to the context model's underperformance, with obfuscation being one of the reasons.
We further conducted a study to investigate the reasons behind the underperformance of PassFinder.
Details of this study are discussed in Section~\ref{ssec:passfindereval}}.

We further categorized all the detected \apikeys of each tool to the corresponding cloud providers, as shown in Table \ref{tab:filter}.
Among all the cloud service providers that are supported by each tool, only a few are detected in our dataset. \revision{ 
99.9\% of the detected \apikeys are from Google Cloud service and Twitter Cloud service.} Additionally, LeakScope was able to detect a more significant number of Google \apikeys compared to the Three-Layer Filter. 

For Twitter Client ID/Secret, the Three-Layer Filter identified 88,172, whereas Leak-Scope only identified 10. \revision{This large number of detections from the Three-Layer Filter is expected, as Twitter credentials themselves can take the form as Twitter credentials themselves can take the form as random alphanumeric strings of lengths ranging from 18–25 or 40–50 characters. To capture this property, only relatively loose regular expressions can be used, which unavoidably match a wide range of random strings.} Upon analyzing a sample of the 88,172 detected strings, we found that although they exhibit the structural characteristics of API keys, none of them were actual Twitter credentials. We further discuss the implications of this observation in Section~\ref{ssec:fp}.

\textbf{Summary:} Overall, there are more \apikeys that the Three-Layer Filter can identify compared to other tools, demonstrating its ability to detect random secret-like strings. However, this result alone does not demonstrate its effectiveness. Evaluating false positives and negatives is equally critical in assessing the tool's effectiveness.

\subsection{RQ2.2-False Positive}\label{ssec:fp}
\subsubsection{RQ2.2-Methodology}\label{sssec: 2_2_M}
In this section, we aim to compare the false positives reported by each tool. As described in Section \ref{sssec:ethics}, it is impractical and impossible to validate all the obtained \apikeys fully. Consequently, we conducted a manual analysis to identify the false positives.
Furthermore, since only \apikeys belonging to Google API key and Twitter client ID/Secret are detected by at least two tools, we only selected these \apikeys for the comparison of false positives. Additionally, PassFinder is excluded from this comparison as it did not detect any \apikeys. A further experiment to comprehensively evaluate PassFinder is detailed in Section \ref{ssec:passfindereval}.

However, since the Three-Layer Filter detected 88,172 Twitter client IDs/Secrets, manually validating all of these \apikeys was unfeasible. To address this, we randomly sampled 383 \apikeys from the 88,172, ensuring a 95\% confidence level with a 5\% margin of error~\cite{randomsample}. Then, we manually analyzed the randomly sampled Twitter client IDs/Secrets and all the Google API keys to identify any false positives as described in Section \ref{sssec:ethics}.
\subsubsection{RQ2.2-Result}
\revision{After manually analyzing a total of 2,117 Google API keys (1,816 detected by LeakScope and 664 by the Three-Layer Filter) and 393 Twitter secrets (10 from LeakScope and 383 from the Three-Layer Filter)}, we discovered that for Google, neither LeakScope nor the Three-Layer Filter falsely detected \apikeys. Their precision all reached 100\%.
In addition, both selected tools consider the reduction of false positives to be one of their main objectives.
Three-Layer Filter leverages entropy filter, word filter, and pattern filter to reduce false positives.
LeakScope identifies \apikeys via backward slicing starting from the arguments of cloud APIs, ensuring that the extracted strings are \apikeys.

However, we made different observations for the Twitter client ID/Secret. The 10 Twitter \apikeys detected by LeakScope are all considered to be valid Twitter \apikeys through manual analysis. In contrast, none of the 383 Twitter \apikeys identified by the Three-Layer Filter were considered valid Twitter \apikeys.
\revision{
Upon closer inspection, these strings appeared random but generally fell into two categories: (1) 68\% are tokens associated with other services such as Facebook or Crashlytics, and (2) 32\% are benign random strings unrelated to authentication, such as session identifiers or serial numbers. This high false positive rate stems from the use of loose regular expressions for Twitter in the Three-Layer Filter, which capture any random string of a given length.

These findings highlight a core limitation of the Three-Layer Filter: while it can distinguish random from ordinary strings, randomness alone is insufficient to determine whether a string represents a valid secret. More broadly, this illustrates the weakness of methods that rely only on intrinsic string characteristics, as they tend to overgeneralize and misclassify benign data when consistent and recognizable patterns are absent.}

\textbf{Exploitable Checked-in Secrets.}
We acknowledge that our manual analysis may be subject to biases. Our manual analysis process may falsely consider a string to be a valid \apikey or vice versa. To address this concern, we validated the Google API keys by sending requests.
We implement this validation process to demonstrate the exploitability of the detected \apikeys.
After validating all the obtained Google API keys by sending a request to the server, 74 Google API keys granted access to the Google Map service. We reported these Google API keys to the developers of the corresponding \androidapps, and we received responses from 13 \androidapps that acknowledged the issue. This validation signifies that most of the \apikeys detected in this study are likely to be exploitable, revealing the severity of the \apikeys leakage in Android apps.

\textbf{Summary: } Overall, only 2,142 detected \apikeys are certain to be valid after manual validation. Among the tools, LeakScope demonstrated its strength in reducing false positives, as it produced none for both Google API keys and Twitter Client IDs/Secrets. In contrast, while the Three-Layer Filter produced no false positives for Google API keys, it generated a significant number for Twitter Client IDs/Secrets. This indicates that the Three-Layer Filter's filtering process is less effective when handling loose regular expressions.
\subsection{RQ2.3-False Negative}\label{sssec:fn}
\subsubsection{RQ2.3-Methodology}\label{sssec: 2_3_M}
In this section, we aim to compare the false negatives of each tool.
It is impractical to obtain the ground truth of our dataset as it requires manual labelling of over 1 million strings.
In addition, all these strings are used in obfuscated code with minimal meaningful contexts.
As a result, we cannot calculate the exact recall of each tool.
To address this problem, we evaluate the overlap between the detection results of the LeakScope and the Three-Layer Filter. PassFinder was not included as it did not detect any \apikeys.

Furthermore, as described in the Section \ref{sssec:dc}. We could not equalize the detection capability of LeakScope and the Three-Layer Filter. 
As a result, we focused solely on \apikeys categorized as Google API keys for a fair comparison, since both of the tools detected a significant number of valid Google API keys. Additionally, to ensure fairness, we removed duplicate Google API keys detected within the same tool, resulting in a final set of 312 unique keys for the Three-Layer Filter. Twitter client ID and secrets are excluded because the Three-Layer Filter failed to detect any valid Twitter client and secrets after manual validation.
\subsubsection{RQ2.3-Result}
\begin{figure}[t]
\caption{Venn Diagram for overlap of the number of unique detected Google API keys for LeakScope and Three-Layer Filter.}
\centering
\includegraphics[width=0.5\linewidth]{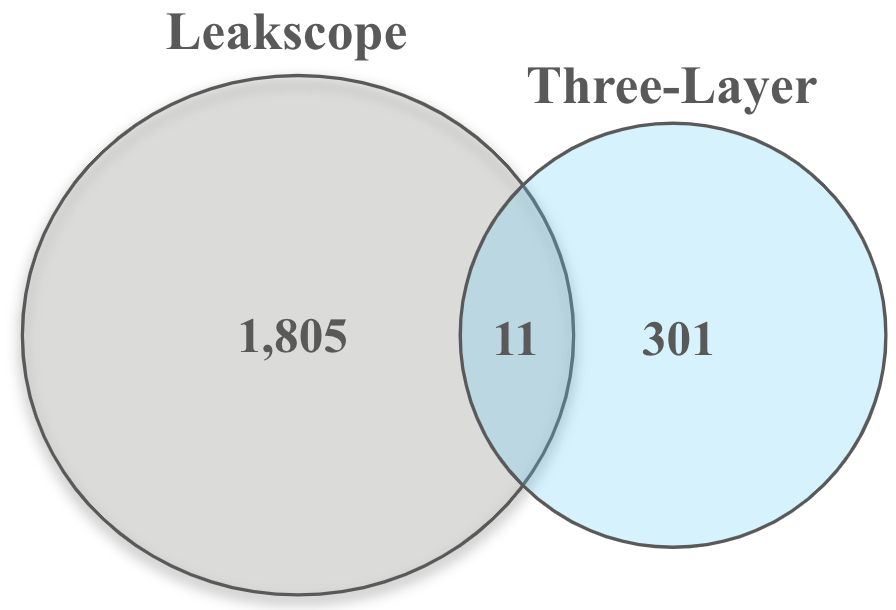}

\label{ven_d}
\end{figure}

The Venn Diagram in Figure \ref{ven_d} shows the overlap of the detected \apikeys between each tool. The Three-Layer Filter identified 374 Google API keys, 11 of which were also detected by LeakScope. Conversely, LeakScope identified 1,816 Google API keys, with 11 overlapping with those found by the Three-Layer Filter. The minimal overlap between the tools suggests that both of them have missed detecting a significant number of \apikeys. Even with the absence of the ground truth, this signifies that all the tools have a relatively low recall. However, based on the overlap, we can still conclude that LeakScope missed fewer \apikeys compared to the Three-Layer Filter. 

We further investigated the \apikeys that can only be detected by LeakScope or the Three-Layer Filter. We found that the difference between detection strategies is the main cause of the minimal overlap.
For LeakScope, it can extract the strings that are not complete in the Android apps. For example, if a \apikey needed to be obtained through several string manipulations such as \textit{StringBuilder.append()} before being used, the Three-Layer Filter would not be able to detect it as the incomplete string can not match the regular expressions.  
In contrast, LeakScope is highly dependent on the pre-defined cloud APIs. However, there exist \apikeys that are encapsulated in the \androidapps but not passed to the cloud APIs. This explains why LeakScope also failed to detect some \apikeys. This result suggests that these tools are rather complementary.

\textbf{Summary: }Overall, although the exact number of false negatives could not be determined, the minimal overlap between the tools suggests that both suffer from a significant issue of producing a large number of false negatives. 

\subsection{A Further Evaluation of PassFinder}\label{ssec:passfindereval}
As demonstrated in Table~\ref{tab:filter}, PassFinder did not detect any \apikeys in this experiment when combining the two models. To thoroughly analyze the performance of PassFinder, we conducted several additional experiments.
\subsubsection{Manual Inspection}
In our experiment, the password model of PassFinder identified 11,343 strings as potential \apikeys. However, the context model concluded that none of these were valid \apikeys. To investigate further, we aimed to determine the validity of these strings and explore the reasons behind the context model's failure to identify any of them.
Therefore, we first randomly sampled 372 strings with their contextual surroundings, achieving a 95\% confidence level with a 5\% margin of error~\cite{randomsample} from the 11,343 strings identified by the password model. Upon manual investigation, we found that it is similar to the Three-Layer Filter with loose regular expressions; although these 372 strings are randomly generated strings, their context did not provide sufficient information to support that they are valid \apikeys. This was primarily due to obfuscation, which significantly differed from the training data used for the context models. The training set consisted of source code extracted from GitHub that contained meaningful, unobfuscated terms. Additionally, the extracted context was presented in Jimple format, an intermediate representation of Java, which further contributed to PassFinder's ineffectiveness.
To evaluate the impact of the Jimple format, we conducted an additional study where the Jimple format was transformed into the Java format. The details of the study are detailed in the following section.
\subsubsection{Jimple to Java Transformation}
The use of the Jimple format instead of Java may have impacted the performance of the context model. Moreover, since the original sample did not include any valid \apikeys, it is challenging to fully determine the effectiveness of the context model. To address this, we created an additional test set containing 664 valid \apikeys from Google Cloud services detected by the Three-Layer Filter, along with their contextual surroundings manually converted into Java format. The conversion process was straightforward, as Jimple is an intermediate representation of Java. Listing \ref{lst:APKJimple} shows a Jimple-formatted code snippet from an Android app, closely resembling Java code and providing sufficient information to reconstruct the corresponding Java code. Listing \ref{lst:APKJAVA} presents the transformed Java-format snippet from Listing \ref{lst:APKJimple}.

\begin{code}
\begin{lstlisting}[language=Java,
    numbers=left,
    numbersep=5pt,
    frame=single,
    tabsize=2,
    caption={Extracted code snippet in Jimple format},
    label={lst:APKJimple}
]
$r0 = newarray (java.lang.String)[3]
$r0[0] = ""
$r0[1] = "67"
$r0[2] = ""
<jp.b: java.lang.String[] c> = $r0
\end{lstlisting}
\end{code}

\begin{code}
\begin{lstlisting}[language=Java,
    numbers=left,
    numbersep=5pt,
    frame=single,
    tabsize=2,
    caption={Transformed Java code snippet from Listing \ref{lst:APKJimple}},
    label={lst:APKJAVA}
]
String[] array = new String[3];
array[0] = "";
array[1] = "67";
array[2] = "";
jp.b.c = array;
\end{lstlisting}
\end{code}

The context model still did not identify any from the test set. After analyzing the Java code snippets from the training set, we found that the majority of the code snippets that contain passwords are the declaration of global variables of classes. In contrast, the code snippets containing the checked-in secrets in APK files are the locations where the checked-in secrets are used, such as the example we discuss in Listing 1. 
This indicates a fundamental difference between the contexts in the training set and those in the test set. Specifically, the training data does not adequately represent the usage patterns of checked-in secrets found in APK files. This misalignment may explain the context model's inability to detect these secrets effectively.

\subsubsection{Password Model on Labelled dataset}
We also conducted an additional experiment to fully validate the performance of the password model, as the random sample strings in the previous section did not contain any valid \apikeys upon manual verification. 
Using the same test set from the previous section. This test aimed to evaluate the model's ability to identify 664 valid \apikeys. However, the password model only identified 62 \apikeys from the test set, underscoring its restricted capability.
The underperformance of the password model may be attributed to the nature of its training set. The machine-generated passwords in the training set were created using a random string generator~\cite{passfinder}, which likely differs from the structure and characteristics of actual \apikeys.

%% file: RQ3.tex
\section{RQ3: Implications}
In RQ3, we discuss the key implications of this research, including the limitations of the existing \apikey detection techniques and potential directions for future research in developing a \apikey detection tool more suitable for \androidapps.
\subsection{Limitations of the Evaluated Tools}
In this section, we summarize the limitations of the evaluated tools.
\subsubsection{Limitation of Three Layer Filter (Intrinsic Value Analysis)}
\ul{\textit{First, the detection capability of the Three-Layer Filter is constrained by the manually crafted regular expressions.}}
Regular expression matching is the most important step in the Three-Layer Filter. The number of manually crafted regular expressions determines the detection capability of the Three-Layer Filter. The strings identified by a precise regular expression will likely not require the following filtering process. In our experiment, only one string was excluded from the strings that were obtained through precise regular expressions. However, acquiring precise regular expressions for all categories of \apikey can be challenging. Some secrets, such as Google API keys, tend to follow recognizable and consistent formats, making them easier to capture with well-defined patterns. In contrast, secrets issued by other providers, such as Twitter Client IDs or Secrets, often consist of arbitrary alphanumeric strings without a stable structure. \revision{For instance, Twitter secrets may appear as random strings like ``xvz1******Bog'' or ``L8qq9PZ******0vZk'', which are difficult to distinguish from non-secret strings using regex alone.} As a result, it is challenging to expand the detection capability of the Three-Layer Filter.

\ul{\textit{Second, valid checked-in secrets can be excluded in the arbitrary filtering steps based on entropies, patterns, and words.}}
The entropy and pattern filter in our experiment excluded ten strings that are very likely to be a potential \apikey through our manual analysis.
The threshold of the pattern sequence is set to be a length of four, which is problematic as some \apikey could contain such a pattern. \revision{A concrete example is Firebase Cloud Messaging API keys, which often begin with four identical characters such as \textquotedblleft AAAA...\textquotedblright. These legitimate keys were mistakenly excluded due to the repetitive pattern.}

Similarly, while the word filter did not lead to any false negatives in our dataset, it poses a risk of filtering out valid keys that happen to include common English words. \revision{For instance, a \apikey stored in a string like \textquotedblleft key=AIzaSyB...\textquotedblright or \textquotedblleft token=abc123..." could be excluded simply because it includes substrings like \textquotedblleft key" or \textquotedblleft token", even though the value is a valid \apikey.} Since randomly generated keys can occasionally contain English words by chance, it's difficult to establish filtering rules that are both strict enough to eliminate noise and flexible enough to avoid excluding true positives. As a result, designing precise and reliable filters remains a significant challenge.

\ul{\textit{Finally, regular expressions cannot identify checked-in secrets that are computed from several strings.}}
After analyzing the \apikeys detected exclusively by LeakScope, but which fall within the detection capability of the Three-Layer Filter (same category), we found that the Three-Layer Filter fails to extract \apikeys that are obtained through string manipulation methods such as \textit{append()}. \revision{For example, if an \apikey is constructed in an \androidapp using multiple string concatenation operations like \textquotedblleft AIza" appended with \textquotedblleft SyA...", the resulting key will not exist in its entirety in the static string pool, making it invisible to regex-based detection.} Regular expressions also rely heavily on the total length of the string to enforce structure, and when keys are split or dynamically assembled at runtime, their fragmented form falls outside the expected pattern length, preventing a successful match. This limitation underscores the need for detection techniques that can handle non-literal and dynamically constructed secrets.

\subsubsection{Limitation of LeakScope (Static Analysis)}
\ul{\textit{The primary limitation of LeakScope can only detect checked-in secrets from a list of pre-defined cloud APIs.}}
To improve the detection capability of LeakScope, significant manual effort is required to expand and maintain the list of cloud API signatures. This dependency on predefined APIs substantially limits its ability to generalize or detect secrets outside the curated set. \revision{For example, during our study, we attempted to identify any cloud APIs that accept PayPal \apikeys as input but were unable to find a relevant API call to include in the signature list. As a result, even if a PayPal API key were present in the code, LeakScope would be unable to detect it due to the absence of a matching API in its predefined list.} This example illustrates the inherent limitation of relying on a fixed set of API signatures for detection.

\ul{\textit{Second, the unsound static analysis of LeakScope fails to recover some values of the checked-in secrets.}}
LeakScope failed to recover the value of 1,353 identified locations of \apikeys.
LeakScope is capable of extracting \apikeys even if they are stored as environment variables in files such as XML or JSON. However, not all environment variables can be extracted. LeakScope uses APKTool to reverse engineer APK files. Due to the limitations of APKTool, it may sometimes fail to obtain the corresponding files that store environment variables from APK files, leading to the failure of \apikey extraction. Moreover, APKTool may also fail to decompile some classes from the APK file, which would affect the performance of LeakScope.

Additionally, since LeakScope uses Soot for backward analysis, it is also subject to the limitations of Soot. Other than the failures caused by the missing files during the reverse engineering process, LeakScope also failed to extract the exact value \apikeys that are stored as static variables within a Java class. e.g., {public static String Google API key = \textquotedblleft Alza****..". Soot sometimes can only get the name of these variables. 

\ul{\textit{Third, the incorrect Cloud API signature generation and matching can impact the accuracy of detection.}} We observed that LeakScope occasionally mismatches the Cloud API in the APK file. Some irrelevant APIs that have the same package structure and similar content would also be considered as one of the pre-defined cloud APIs during the analysis phase of LeakScope. \revision{This limitation arises because LeakScope primarily relies on package structures when generating cloud API signatures. Consequently, methods from unrelated libraries that share similar package hierarchies or naming conventions may be misclassified as valid API calls. For example, we found that the method \texttt{setOAuthConsumerKey()}, which is typically associated with Twitter API usage, was sometimes incorrectly matched to unrelated methods residing in third-party libraries that resemble the structure of the official Twitter SDK. This kind of false matching can lead to inaccurate detection results and highlights the challenge of relying solely on static API signature matching without additional semantic validation.}

\ul{\textit{Finally, there exist checked-in secrets that are not used by cloud APIs.}} After analyzing the \apikeys that were only detected by the Three-Layer Filter, we observed that these \apikeys were not used as parameters for any of our selected cloud APIs. Some developers prefer to send requests directly through URLs or use other functions that we are unaware of. As the Venn diagram of our result in Figure \ref{ven_d} indicates, this is not an uncommon situation.

\subsubsection{Limitation of PassFinder (Machine-Learning-Based Technique)}
When combining the predictions made by both the context model and the password model, PassFinder did not detect any \apikeys.
To understand the reason behind this, we analyzed the two models separately.

\ul{\textit{We observed that the poor performance of PassFinder can be attributed to (1) the training set and (2) the context modelling under obfuscation and insufficient information from 12 lines of context.}}

For the password model, the provided training set is a combination of ordinary strings, human-chosen passwords, and machine-generated passwords. These strings may differ from \apikeys, although they all appear random.
Checked-in secrets of the same cloud service often adhere to specific patterns.
As presented in Section \ref{ssec:passfindereval}, this difference in characteristics may be the reason why the password model only detected 62 out of 664 \apikeys in the training dataset.

For the context model, even though we transformed code snippets from Jimple format to Java format within the training set, the effectiveness of these snippets was diminished by obfuscation. This process replaced variable, method, and class names with random, meaningless letter sequences, reducing their overall significance. Furthermore, we observed that simply 12 lines of code could not always give enough information to detect the presence of \apikeys.

\revisionM{\textbf{Summary}: In summary, each analyzed method exhibits intrinsic limitations that are not specific to Android but are fundamental to their underlying detection methodologies. However, the complexity of Android APK structures and the pervasive use of obfuscation introduce additional challenges that further constrain their performance.
The Three-Layer Filter is limited by its reliance on manually defined regular expression rules and its assumption that secrets appear as complete string literals in source code. These constraints prevent it from recognizing secrets that are dynamically constructed, encoded, or transformed at runtime. In the Android environment, the effectiveness of this approach can be further reduced by obfuscation or string encoding, which alter literal values or conceal them from direct inspection. In some cases, missing or corrupted files during decompilation may also lead to incomplete coverage of code units, further limiting the Three-Layer Filter’s detection scope.
LeakScope, while more sophisticated in design, is strongly affected by the characteristics of the Android environment. Its reliance on static analysis and predefined API signatures restricts detection to known APIs and simple invocation structures. In practice, the event-driven architecture of Android applications, combined with the complexity of component interactions and the widespread use of code obfuscation
further hinder data-flow propagation. For instance, among 3,179 identified cloud API usages, only 1,846 corresponding values were successfully extracted. Additionally, we observed five cases where incorrect APIs were used for further data flow analysis due to mismatches caused by code obfuscation.
Quantitatively, the overlap between the Three-Layer Filter and LeakScope detections among all identified Google API keys is only 0.5\% as shown in Figure \ref{ven_d}, indicating that these deterministic methods capture largely distinct subsets of secrets and that many valid cases remain undetected because of their differing assumptions.
Machine-learning-based approaches, such as PassFinder, provide greater flexibility by learning diverse representations of secret usage rather than relying on predefined patterns. Nevertheless, pervasive obfuscation in production Android apps significantly degrades their performance by altering or removing contextual cues essential for effective learning. These challenges call the need for Android-aware learning strategies, which motivated our preliminary work on String Group Analysis, a framework designed to enhance the robustness of learning-based detection under obfuscation.}
\subsection{Future Direction}
\subsubsection{Complimentary Results of the Tools}
As described in Section~\ref{sssec:fn}, all the tools missed detecting a significant number of \apikeys, indicating that the actual recall of each tool is low.
\revision{The minimal overlap between the Three-Layer Filter and LeakScope stems from their fundamentally different detection strategies. The Three-Layer Filter targets complete strings stored directly in the code, making it effective for identifying hardcoded, static secrets. In contrast, LeakScope applies inter-procedural analysis with control and data-flow reasoning, which enables it to capture secrets that are dynamically constructed or spread across program elements. Consequently, LeakScope can recover a broader range of string values.

Despite these strengths, both approaches have inherent limitations that restrict their detection capabilities. The Three-Layer Filter is constrained by the coverage of its regular expressions, while LeakScope relies on a predefined set of cloud API signatures. While effective for cloud-related secrets, this reliance risks overlooking other formats, such as secrets embedded in custom requests or transmitted through non-standard APIs. Moreover, both regular expressions and cloud API specifications are difficult to comprehensively enumerate and formalize.

These trade-offs highlight the complementary nature of the two techniques. LeakScope’s data-flow analysis can help reconstruct or complete string values that are otherwise inaccessible to regex-based matching, thereby enhancing the effectiveness of the Three-Layer Filter. At the same time, regex-based approaches can efficiently detect fully formed strings that may not be directly associated with known API usage. Together, such methods can offer improved recall by mitigating each other’s blind spots.}

    \revisionM{
This complementary relationship is also reflected in the quantitative results. Although calculating exact false-negative rates is infeasible due to the absence of a ground-truth benchmark for checked-in secrets in Android apps, additional quantitative analysis was conducted to better illustrate the complementarity between LeakScope and the Three-Layer Filter. Among all valid Google API keys detected, only 11 were found by both tools as shown in Figure \ref{ven_d}, which is a minimal overlap account for 0.5\% of all detected Google API keys. This indicates that they capture largely distinct subsets of secrets. In particular, 99.4\% of LeakScope’s detections and 97.1\% of the Three-Layer Filter’s detections are non-overlapping, and combining the two approaches increases overall detection coverage by roughly 20\% compared with using LeakScope alone. These quantitative results reinforce that integrating syntactic- and flow-based analyses yields broader coverage and complementary detection capabilities.}

In contrast, machine-learning-based tools, in general, can adapt to a wider range of cases since they do not require any pre-defined patterns. However, our results suggest that simply providing more training sets cannot yield a significant improvement. The effectiveness of the context model is significantly degraded by the presence of obfuscation. Relying on code snippets as input for the model is proven to be ineffective in our evaluation. This finding suggests the need for a new feature extraction.

\subsubsection{Strings as Features for Secret Detection}
\begin{table}[t]
\caption{String Group that contains a Google API key}
\centering
\begin{tabular}{ll}
\toprule 

String Group ID & Strings\\ \midrule

String Group 1 & grant\_type, refresh\_token, client\_id, client\_secret, refresh\_token \\ 

String Group 2 & account, google, user\_id, client\_secret, content\\ 

String Group 3 & builder, google, refresh, token, builder \\

\bottomrule

\end{tabular}

\label{tab:stringGroup}
\end{table}
After analyzing the methods that contain a \apikey, we propose to use \textbf{string groups} as features to detect \apikeys. A string group refers to all the strings within one single method. In our manual inspection, we observed that string groups that contain \apikeys shared some similar patterns. For instance, the names of cloud providers like \textquotedblleft Google\textquotedblright or \textquotedblleft Firebase \textquotedblright, along with terms related to secrets such as \textquotedblleft user\textquotedblright,  \textquotedblleft token\textquotedblright, or \textquotedblleft key\textquotedblright, frequently appear in these groups. As an example, Table \ref{tab:stringGroup} is three string groups that were extracted from different \androidapps which contain a Google API key. While the individual strings within these groups are not necessarily identical, they exhibit notable similarity due to the frequent occurrence of certain words (e.g., refresh, token, client\_id, etc.).

This suggests that the patterns in string groups could serve as potential indicators of \apikeys. Moreover, since string literals are typically not modified by the obfuscation of apps in Google Play Store~\cite{obfuscation}, it can resolve the existing challenge caused by obfuscation for \apikey detection in Android apps. We conducted two preliminary studies to test this potential research direction.

\noindent\textbf{Study 1: Statistical Hypothesis Testing}
To verify the feasibility of this approach, we first conducted a preliminary study with the hypothesis that string groups containing checked-in secrets exhibit similarities to one another and are distinct from those that do not contain any secrets. To test this hypothesis, we conducted a statistical analysis. First, we collected two collections of string groups: \textit{Collection Secret} included 602 string groups containing Google API keys and secrets obtained by the Three-Layer Filter. we excluded string groups that contain only the \apikey without any additional strings. To prevent potential bias in terms of quantity, the second set, \textit{Collection NoSecret}, included 602 randomly selected string groups that did not contain any checked-in secrets. 
Methods with only a single string were excluded to ensure meaningful comparison. 
We then calculated the similarity between string groups using cosine similarity~\cite{cos}, which is better for this task as it works with non-English text, unlike Levenshtein distance~\cite{lev} or Word Movers’ Distance~\cite{wmd}. Cosine similarity measures the cosine of the angle between two character vectors, generating a value between 0 and 1, where 1 indicates identical word vectors and 0 indicates no similarity. The similarity score is calculated using the equation below~\cite{cos}.
\begin{equation}
   \text{sim}(x, y) = \frac{x \cdot y}{\|x\| \cdot \|y\|}
\end{equation}

For each string group in Collection Secret, we first transform the string groups into the character vector, in which the number in the vector represents the frequency of a character.
Then, we computed the average cosine similarity both within Collection Secret (Comparison-SvS) and between Collection Secret and Collection NoSecret (Comparison-SvN).
We need to verify that the two comparisons are statistically different to test our hypothesis. We applied both F-test~\cite{ftest} and Z-test~\cite{ztest}. These tests assessed whether the means and variances of the cosine similarity scores in Comparison-SvS and Comparison-SvN were significantly different. The statistical significance was determined through p-values, with smaller p-values indicating a greater likelihood that the differences in means or variances were not due to chance. Additionally, given the various forms that strings can take in Android apps, we modified Collection Secret and Collection NoSecret into three distinct versions: case-sensitive, case-insensitive, and English Word Extraction. The first two variations considered whether string case impacted the similarity calculations, while the English Word Extraction version isolated all English words using a dictionary~\cite{englishdict} for further analysis.

The result of the two tests is shown in Table \ref{tab:statT}. The calculated p values of the two tests for the three versions are all very close to 0 and much smaller than the common significance level alpha (0.01)~\cite{alpha}, indicating that differences of Comparison S-S and Comparison S-N are statistically significant. We observed that the difference in the case-sensitive comparison is larger than in other variations. This is likely because ordinary strings tend to use more capital letters.
Furthermore, the mean similarity score in Comparison S-S is also higher than in Comparison S-N. 

Such test results demonstrate the potential to leverage string groups to detect \apikeys in obfuscated Android apps. We plan to further explore this direction to propose a new Machine-Learning-Based \apikey detection technique based on string groups.

\begin{table*}[t]
\centering
\caption{Result of F-test and Z-test}
\label{tab:statT}
\begin{tabular}{ccccl}
\toprule
                        & \textbf{F-Test(p value)} & \textbf{Z-Test(p value)} & \makecell{\textbf{Similarity Score} \\ (Comparison s-s:\\ Comparison s-n)} &  \\ \midrule
Case-Sensitive          & $1.76 \times 10^{-14}$
               & 0               & 0.55 : 0.13                                           &  \\
Case-Insensitive        & $7 \times 10^{-13}$               & 0               & 0.65 : 0.54                                           &  \\
English Word & $7.6 \times 10^{-12}$               & 0               & 0.65 : 0.55                                           &  \\ \bottomrule
\end{tabular}
\end{table*}

\noindent\textbf{Study 2: Binary Classifiers Using String Groups for Checked-in Secret Detection}

To demonstrate the potential of string group analysis, we conducted an additional experiment to actually train machine learning models with string groups. 

We first extracted features from the \textit{Collection Secret} and \textit{Collection NoSecret} datasets using Term Frequency-Inverse Document Frequency (TF-IDF) and Count Frequency, respectively. These extracted features were then used to train three classification models: Logistic Regression, Naive Bayes, and Support Vector Classification (SVC). 
Using this experiment, we aim to evaluate the performance of using string groups as features to detect \apikeys.

\ul{\textit{\textbf{Dataset Pre-processing}}} We used the same dataset as in Study 1, and similar to statistical hypothesis testing, we transformed the two collections into three distinct versions: case-sensitive, case-insensitive, and English word extraction. The case-sensitive and case-insensitive
versions examined the impact of string cases. The English word
extraction version focused on isolating all English words from the string groups using a dictionary~\cite{englishdict}. 
Each version was then used to generate collection pairs of the same type, which were combined to construct the complete dataset. As a result, we obtained three distinct datasets, each corresponding to one of the three versions.

\ul{\textit{\textbf{Feature Extraction}}} We selected two simple and relatively popular feature extraction techniques used in the domain of text classification: TF-IDF and Count frequency. 

\textbf{TF-IDF} is a statistical measure used to evaluate the importance of a word in a document relative to a collection of documents (string groups). It multiplies two following metrics~\cite{tfidf}.
\begin{itemize}
    \item Term Frequency (TF): This measures how frequently a term appears in a document.  In our case, each string group is treated as a document. It is computed by counting the occurrences of a term within a string group, as shown in Equation \ref{eq:tf}.
    \begin{equation}
    tf_{\textit{t,d}} = \text{count(t,d)}
    \label{eq:tf}
    \end{equation}
    \item Inverse Document Frequency (IDF): This assesses the importance of a term within the entire corpus. It is calculated by taking the logarithm of the total number of documents(string groups) divided by the number of documents(string groups) containing the term, as shown in Equation \ref{eq:idf}.
    \begin{equation}
        idf_{\textit{t,d}}=\log_{10}\dfrac{\text{Total Number of documents}}{\text{Total Number of documents that contains the term t}}
        \label{eq:idf}
    \end{equation}
    
\end{itemize}

\textbf{Count Frequency} simply counts the occurrences of each word in a document(string group) without considering their distribution across the whole dataset.

Both feature extraction methods produce a vector representing the computed value of each term for every string group.

\ul{\textit{\textbf{Model Training}}} We first split the datasets into an 80\% training set and a 20\% testing set. Using the scikit-learn package~\cite{log,svc,naive}, we trained three models (Logistic Regression, Naive Bayes, and SVC) on each dataset separately. For each model, we selected the hyperparameters that achieved the highest training accuracy.

\ul{\textit{\textbf{Result}}} We calculated the precision, recall, accuracy and F1 score  for each selected model on the different datasets. These metrics were used to evaluate the performance of the models, providing a comprehensive understanding of their effectiveness across the various versions of the datasets. The results of the experiment are listed in Table \ref{tab:modelCS}, Table \ref{tab:modelCI}, and Table \ref{tab:modelEX}, respectively. Each table represents the performance of three models on their corresponding dataset. As shown in the tables, all models demonstrated strong performance, with SVC achieving the highest overall performance, reflected by an average F1-score of 0.97 and an average accuracy of 0.98. Additionally, the two feature extraction techniques evaluated in this study achieved comparable performance, with count frequency performing slightly better than TF-IDF. This is reflected by an average accuracy of 0.96 for both techniques, and average F1-scores of 0.94 for count frequency compared to 0.93 for TF-IDF. However, we observed that the English word extraction version underperformed, which may be due to the fact that some important information, such as URLs and token prefixes, could be ignored during the extraction process.
The results highlight that, despite relying on straightforward feature extraction methods and simple models, string group analysis achieves remarkably effective detection performance. This demonstrates the potential of our approach as a lightweight yet powerful solution.

\revision{While the results demonstrate the effectiveness of the proposed approach, they are based on a small-scale dataset only composed of Google API keys, which may limit the generalizability and robustness of the findings. This controlled setting was chosen due to the availability of sufficient labeled samples for meaningful training and evaluation. To address this limitation, future work will focus on expanding the dataset to include a broader range of secret types and more diverse application contexts.} Furthermore, as discussed previously, the underperformance of the English word extraction version may be attributed to the loss of important information during the pre-processing step. Future work should also focus on developing more advanced pre-processing techniques capable of accurately preserving such information, as well as exploring more complex machine learning models that can better capture these patterns.

\begin{table}[h]
\caption{Result of Three Models on Case Sensitive Dataset }

\label{tab:modelCS}
\resizebox{\columnwidth}{!}{%
\begin{tabular}{l|l|l|l|l}
\toprule
\textbf{}                                                                       & \begin{tabular}[c]{@{}l@{}}\textbf{Precision}\\ (TF-IDF\\ Count Frequency)\end{tabular} & \begin{tabular}[c]{@{}l@{}}\textbf{Recall}\\ (TF-IDF\\ Count Frequency)\end{tabular} & \begin{tabular}[c]{@{}l@{}}\textbf{Accuracy}\\ (TF-IDF\\ Count Frequency)\end{tabular} & \begin{tabular}[c]{@{}l@{}}\textbf{F1-Score}\\ (TF-IDF\\ Count Frequency)\end{tabular} \\ \hline
\multirow{2}{*}{\begin{tabular}[c]{@{}l@{}}\textbf{Logistic} \\ \textbf{Regression}\end{tabular}} & 1                                                                              & 0.99                                                                        & 0.99                                                                          & 0.99                                                                          \\ \cline{2-5} 
                                                                                & 1                                                                              & 0.98                                                                        & 0.99                                                                          & 0.99                                                                          \\ \hline
\multirow{2}{*}{\textbf{Naive Bayes}}                                                    & 1                                                                              & 0.94                                                                        & 0.99                                                                          & 0.97                                                                          \\ \cline{2-5} 
                                                                                & 1                                                                              & 0.94                                                                        & 0.99                                                                          & 0.97                                                                          \\ \hline
\multirow{2}{*}{\textbf{SVC}}                                                            & 1                                                                              & 1                                                                           & 1                                                                             & 1                                                                             \\ \cline{2-5} 
                                                                                & 1                                                                              & 1                                                                           & 1                                                                             & 1                                                                             \\ \bottomrule
\end{tabular}}
\end{table}

\begin{table}[]
\caption{Result of Three Models on Case Insensitive Dataset}
\label{tab:modelCI}
\resizebox{\columnwidth}{!}{%
\begin{tabular}{l|l|l|l|l}
\toprule
\textbf{}                                                                       & \begin{tabular}[c]{@{}l@{}}\textbf{Precision}\\ (TF-IDF\\ Count Frequency)\end{tabular} & \begin{tabular}[c]{@{}l@{}}\textbf{Recall}\\ (TF-IDF\\ Count Frequency)\end{tabular} & \begin{tabular}[c]{@{}l@{}}\textbf{Accuracy}\\ (TF-IDF\\ Count Frequency)\end{tabular} & \begin{tabular}[c]{@{}l@{}}\textbf{F1-Score}\\ (TF-IDF\\ Count Frequency)\end{tabular} \\ \hline
\multirow{2}{*}{\begin{tabular}[c]{@{}l@{}}\textbf{Logistic} \\ \textbf{Regression}\end{tabular}} & 1                                                                              & 0.73                                                                        & 0.96                                                                          & 0.84                                                                          \\ \cline{2-5} 
                                                                                & 1                                                                              & 0.99                                                                        & 0.99                                                                          & 0.99                                                                          \\ \hline
\multirow{2}{*}{\textbf{Naive Bayes}}                                                    & 1                                                                              & 0.94                                                                        & 0.99                                                                          & 0.97                                                                          \\ \cline{2-5} 
                                                                                & 1                                                                              & 0.94                                                                        & 0.99                                                                          & 0.97                                                                          \\ \hline
\multirow{2}{*}{\textbf{SVC}}                                                            & 1                                                                              & 1                                                                           & 1                                                                             & 1                                                                             \\ \cline{2-5} 
                                                                                & 1                                                                              & 1                                                                           & 1                                                                             & 1                                                                             \\ \bottomrule
\end{tabular}}
\end{table}

\begin{table}[]
\caption{Result of Three Models on English Word Extraction Dataset}
\label{tab:modelEX}
\resizebox{\columnwidth}{!}{%
\begin{tabular}{l|l|l|l|l}
\toprule
                                                                 & \begin{tabular}[c]{@{}l@{}}\textbf{Precision}\\ (TF-IDF\\ Count Frequency)\end{tabular} & \begin{tabular}[c]{@{}l@{}}\textbf{Recall}\\ (TF-IDF\\ Count Frequency)\end{tabular} & \begin{tabular}[c]{@{}l@{}}\textbf{Accuracy}\\ (TF-IDF\\ Count Frequency)\end{tabular} & \begin{tabular}[c]{@{}l@{}}\textbf{F1-Score}\\ (TF-IDF\\ Count Frequency)\end{tabular} \\ \hline
\multirow{2}{*}{\begin{tabular}[c]{@{}l@{}}\textbf{Logistic} \\ \textbf{Regression}\end{tabular}} & 1                                                                              & 0.77                                                                        & 0.9                                                                           & 0.87                                                                          \\ \cline{2-5} 
                                                                                & 0.91                                                                           & 0.86                                                                        & 0.9                                                                           & 0.88                                                                          \\ \hline
\multirow{2}{*}{\textbf{Naive Bayes}}                                                    & 0.87                                                                           & 0.91                                                                        & 0.9                                                                           & 0.89                                                                          \\ \cline{2-5} 
                                                                                & 0.8                                                                            & 0.92                                                                        & 0.86                                                                          & 0.86                                                                          \\ \hline
\multirow{2}{*}{\textbf{SVC}}                                                            & 1                                                                              & 0.85                                                                        & 0.93                                                                          & 0.92                                                                          \\ \cline{2-5} 
                                                                                & 0.97                                                                           & 0.85                                                                        & 0.93                                                                          & 0.91                                                                          \\ \bottomrule
\end{tabular}}
\end{table}

%% file: discussion.tex
\section{Discussion}
In this section, we would like to discuss some additional findings and contributions from our research. 

\textbf{Severity of Checked-in Secret Issue in \androidapps.} \revision{Our main study on top-ranked applications demonstrated that \apikey leakage is a persistent and non-trivial security concern. To examine whether these findings also apply to long-tail applications, we conducted an additional experiment on a random sample of 386 applications drawn from the 1,565,139 total reported by AppBrain~\cite{appBrain}, achieving a 95\% confidence level. The experimental setup was identical to the main study, with LeakScope and the Three-Layer Filter configured to the same set of cloud providers. Manual validation was performed for all detected instances across all tools, following the same procedure as in the original experiment.

In the long-tail sample, LeakScope detected 25 \apikeys, consisting of 20 Google API keys, 1 Amazon key, 1 Twitter key, and 2 Facebook keys, all confirmed as true positives. The Three-Layer Filter detected 14 Google API keys, also validated as true positives, but none of the 381 sampled Twitter results were validated as true positives. PassFinder, when applied to the Google API keys detected by the Three-Layer Filter along with their contextual surroundings, did not identify any as valid. 

We further analyzed the sampled applications using their download counts and last update dates, as illustrated in Figure \ref{fig:download} and \ref{fig:updatetime}. The distribution indicates that a substantial portion of the sample that contains leaked \apikeys falls into lower-download categories: 16.67\% have fewer than 1,000 installs, 3.33\% between 1,000–10,000 installs, and 20.00\% between 10,000–100,000 installs, collectively representing over 40\% of the sample. Similarly, a majority of the apps were updated recently, showing that secret leakage is not confined to abandoned or outdated apps. Although these results highlight the presence of leakage in long-tail and less frequently downloaded applications, the additional experiment does not necessarily prove that such apps are inherently more dangerous than popular ones. As demonstrated in our original experiment, widely used applications also suffer from secret leakage, sometimes at significant scale. Taken together, these findings suggest that leakage is not restricted to any single segment of the ecosystem: it affects both niche and highly popular apps, including those actively maintained, underscoring that checked-in secret leakage is a pervasive and systemic problem across Android applications.
\begin{figure}
    \centering
    \includegraphics[width=0.8\linewidth]{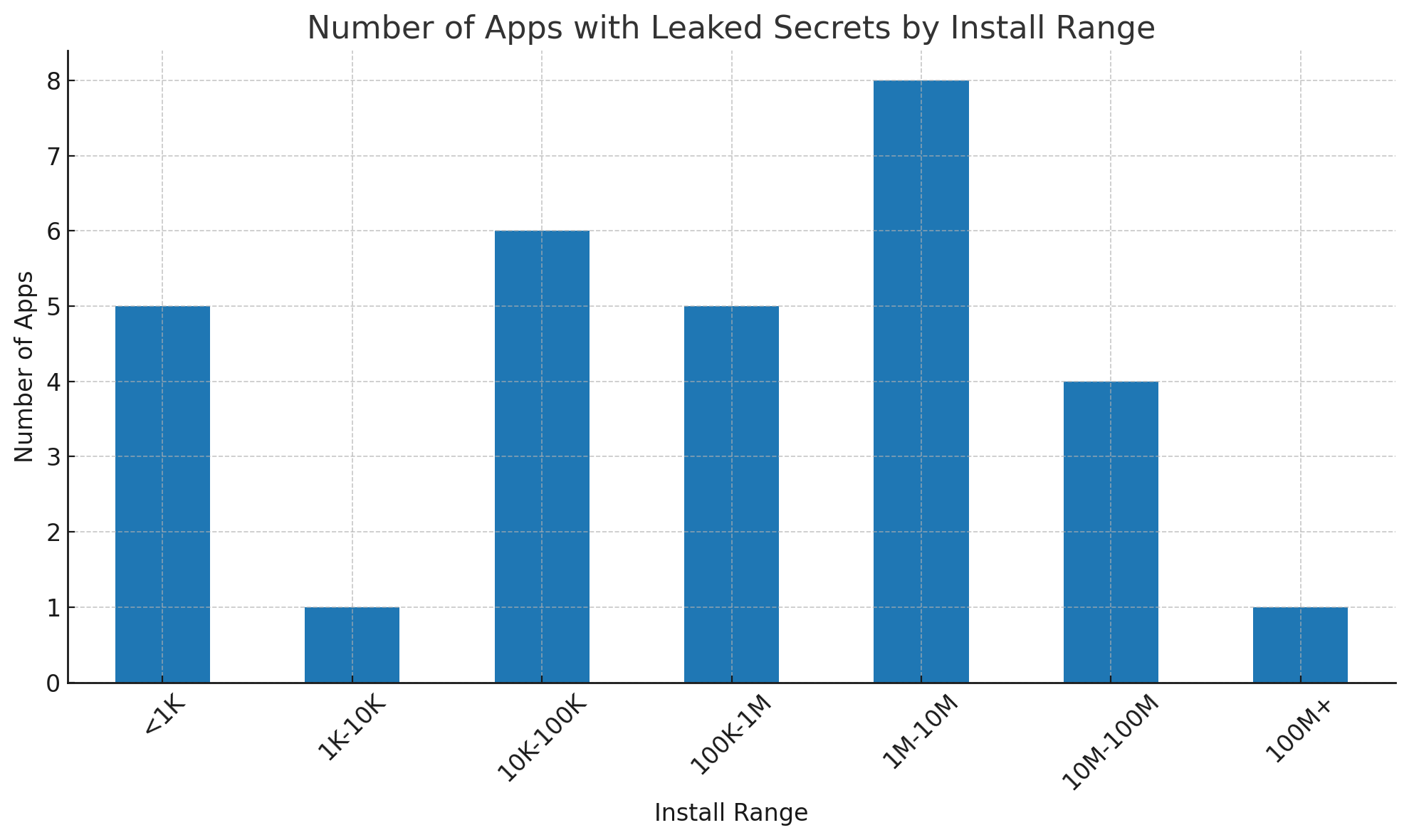}
    \caption{Number of Apps with Leaked Secrets by Install Range}
    \label{fig:download}
\end{figure}

\begin{figure}
    \centering
    \includegraphics[width=0.8\linewidth]{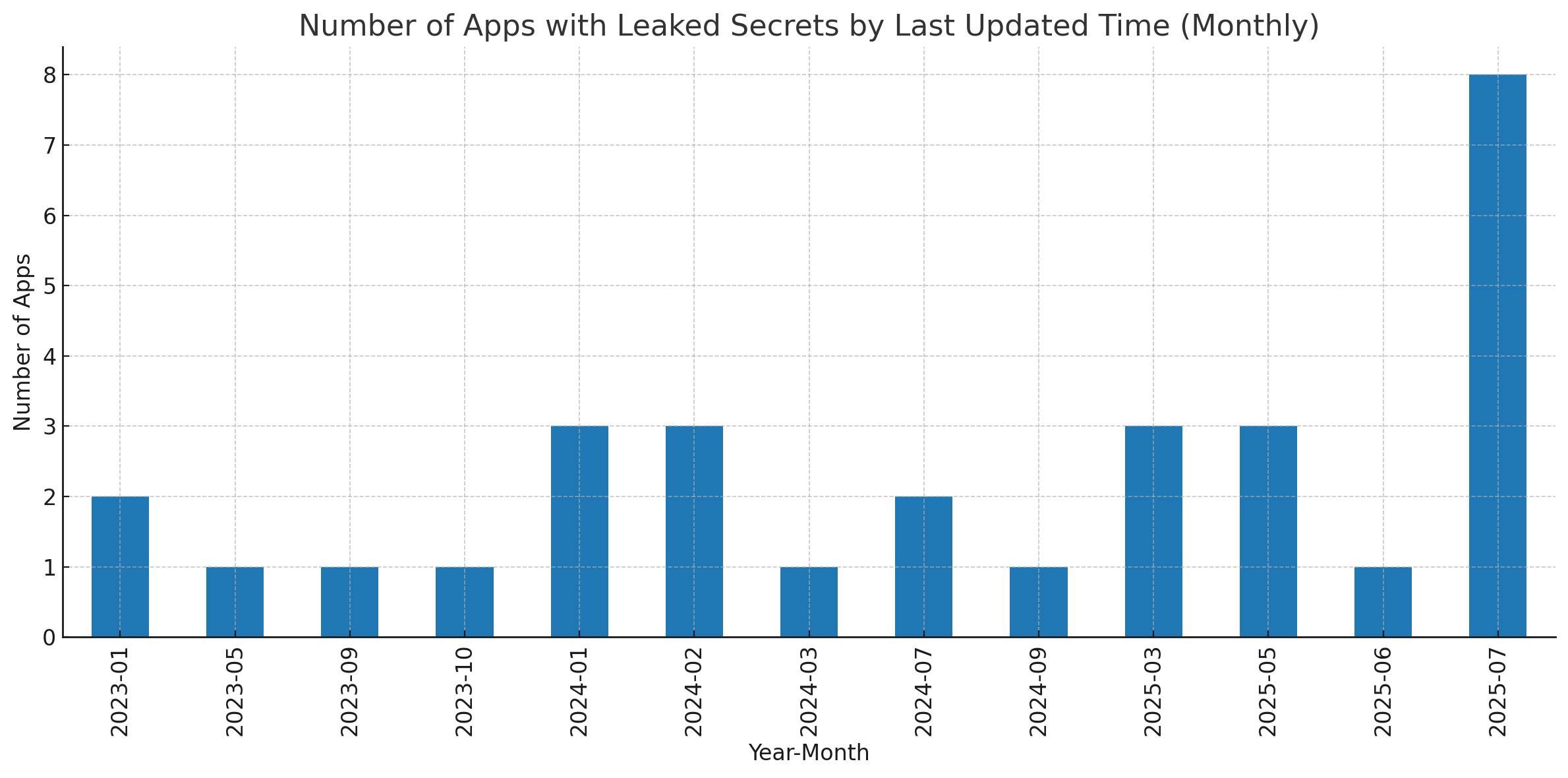}
    \caption{Number of Apps with Leaked Secrets by Last Updated Time (Monthly)}
    \label{fig:updatetime}
\end{figure}

The same limitations observed in the main study were also present in the long-tail sample. High false negative rates and a strong dependency on pre-defined patterns or API lists remain major challenges. This indicates that the risk of \apikey leakage is not limited to top-ranked applications and may be equally prevalent across the broader application ecosystem. Although we could not conduct a larger-scale experiment to precisely measure the relationship between application popularity and security risk due to time constraints, this is an important direction for future research.}

\textbf{Negligence of Checked-in Secret Issue.}
We reached out to the developers of \androidapps affected by the \apikey issue. Some developers acknowledged the problem and committed to addressing it, while others acknowledged the issue but asserted that the \apikey was safe to expose. While it is true that some \apikeys, such as the Firebase API key, are considered safe to expose in code (though not recommended)~\cite{firebase}. Or Twitter Client Secret, which would only become harmful if the associated Twitter Client ID is also leaked, these keys can still pose a security threat when other related information is also exposed. For instance, if project IDs and authentication tokens associated with these \apikeys are leaked, they can compromise security. In our experiment, we successfully extracted additional information associated with the Firebase API key that could be harmful when combined, demonstrating the potential risk even for seemingly harmless \apikeys.

\textbf{A benchmark dataset for \apikeys leakage in \androidapps.} 
We have made available the code snippets and the data flow calculated by LeakScope for all the detected \apikeys, facilitating future research on this subject. The \apikeys and the names of the \androidapps are masked due to the sensitivity of the information.
However, we are willing to share the complete dataset with researchers and developers upon request.

%% file: threat.tex
\section{Threats to Validity}
First, as described in section \ref{sssec: 2_2_M} and \ref{sssec: 2_3_M}, the lack of ground truth is a common challenge in \apikeys related work. Consequently, a significant threat to our study lies in evaluating the false positives for each tool. Without an available benchmark, it was impossible to determine the exact number of false positives, leaving manual analysis as the only means to verify results. To address this, our manual analysis of detected \apikeys would consider not only the format of the strings but also their contextual surroundings. We further validated a part of our \apikeys by sending requests to the cloud services, thereby enhancing the rigour of our manual evaluation process. 
Similarly, we were unable to calculate the recall for each tool. To provide a snapshot of the recall, we analyzed the overlap between the detections of each tool.

Another threat is the detection capability of the selected tools. For instance, the Three-Layer Filter relies on the provided regular expressions as the starting point. Although the authors provided 15 distinct regular expressions, they did not cover all the popular cloud providers. To mitigate this threat, we tried to derive more regular expressions that could find a match in our dataset, and overall, we generated two new regular expressions. In future work, we plan to generate more regular expressions to cover a wider range of cloud providers. As for LeakScope, similar to the Three-Layer Filter, cloud API signatures are the analysis's starting point. The cloud APIs provided by the authors only cover three cloud providers; we collected 13 more cloud APIs covering five more cloud providers. While our collected cloud APIs cannot cover all the available cloud providers, they did improve the detection capability of LeakScope.

It is also important to acknowledge that our evaluation focused on only three detection tools. However, the selection of these tools was grounded in a rigorous systematic literature review, ensuring that they represent the state-of-the-art approaches within their respective categories. Specifically, each tool was chosen to reflect a different detection methodology, allowing for a comprehensive comparison across diverse detection strategies.

In the end, the training and testing process of PassFinder is also one threat to our work. The training set provided by the authors to train the context model does not contain code snippets in Jimple format, while our provided code snippets are in Jimple format. To mitigate this threat, we transformed the code in Jimple format into JAVA format to truly test the performance of PassFinder.

%% file: related.tex
\section{Related work}
\textbf{Checked-in Secret Detection for GitHub.}
Leakage of \apikey issue in the code-sharing platform is a long-studied subject. Besides the tools we selected to run the experiment, there exist other papers that also targeted the \apikey issue. Saha et al.~\cite{reducefalse} introduce a machine learning-based detection tool; they designed 20 features for a string to reduce the false positive, and Lounici et al.~\cite{Lounic} examined the performance of various machine learning model when performing \apikey detection tasks. Basak et al.~\cite{assetha} developed a framework to detect secret-asset pairs in open-source repository with a focus on database key. There are also some community \apikey detection tools. As an example, TruffleHog~\cite{trufflehog}, which was originally designed for GitHub, has also been adapted to the \androidapps on the Google Play store. However, the Three-Layer Filter, which utilizes the same methodology, outperforms TruffleHog. Some researchers have also proposed tools that can detect human-generated passwords. Fadi et al. ~\cite{humanpass} utilized neural networks to predict whether a string is a human-generated password. While Lykousas et al.~\cite{tales} examined the developers' traits when choosing passwords. Those studies all aimed to develop an effective \apikey detection tool for open-source projects but did not examine the feasibility of their approaches on \androidapps.

\textbf{Empircal Study on Checked-in Secrets Issue.}
There have also been several researches that conducted empirical studies on \apikey topics. Basak et al.~\cite{emstudy} evaluated the performance of all the available \apikey detection tools on a benchmark dataset~\cite{secretbench} that contains 818 Github repositories. They also discussed developers' challenges about \apikey by analyzing 779 questions on Stack Exchange~\cite{basak2023challenges}.  Whereas Rahman et al.~\cite{rahman} aimed to investigate the phenomenon of developers disregarding warnings issued by \apikey detection tools, often due to the tools' high false positive rates. Furthermore, Krause et al.~\cite{krause2022committed} analyzed the reason why developers would commit \apikeys by interviewing other developers. Naiakshina et al. ~\cite{Naiakshina_2017} studied how developers deal with \apikey storage through an experiment involving 20 developers. 
However, these empirical analyses have primarily focused on open-source projects, which may not accurately represent the situation in \androidapps. We aim to address this limitation in our research.

\textbf{Security Concerns in Mobile Apps.}
The popularity of mobile apps has also drawn researchers' attention. Many studies were conducted to analyze the security issues in mobile apps. PAMDroid~\cite{pamdroid} was developed to investigate the privacy leakage issue caused by the misconfiguration of analytic service. DroidContext~\cite{DROIDCONTEXT} targeted the detection of privacy leakage and is capable of differentiating malicious privacy identification and benign privacy disclosure identification. Bunyakiati et al.~\cite{secretmanage} examined current secret management and handling in the mobile apps development cycle. Whereas Rahat et al. developed OAUTHLINT~\cite{oauthlint} to detect the security mistakes of the \androidapps that use OAuth API. our work further investigated the security issue caused by \apikey. 

%% file: conclusion.tex
\section{Conclusion}
In this paper, we conducted a literature review to categorize the existing \apikey detection techniques, selected three representative tools and performed an empirical analysis with the selected tools on \androidapps after some adaptation.
We evaluated the performance of the tools based on the number of detected \apikeys, false positives and false negatives.
Overall, \totalN unique \apikeys were detected, affecting 2,115 \androidapps. We further performed a systematic comparison among the tools to analyze the limitations of current \apikey detection techniques. Based on our findings, we proposed possible research directions for developing a more effective \apikey detection tool specifically for \androidapps and conducted two preliminary experiments to support the proposed research directions.
The \apikeys resulted from our experiments form a benchmark for \apikey detection in \androidapps.
The sanitized datasets are available on our project website~\cite{data} without leaking any sensitive \apikeys and the complete dataset can be access upon request.